\documentclass{article}
\usepackage{cite}
\usepackage[pdftex]{graphicx}
\usepackage[cmex10]{amsmath}
\usepackage{subfig, color}

\title{Historical Backtesting of Local Volatility Model using  AUD/USD Vanilla Options}

\author{Timothy G. Ling\\
The School of Mathematical Sciences\\
University of Technology Sydney\\
Ultimo NSW, Australia 2007\\
CSIRO Computational Informatics\\
Sydney, Australia\\
Email: Timothy.Ling-1@uts.edu.au
\and
Pavel V. Shevchenko\\
CSIRO Computational Informatics\\
Sydney, Australia\\
The School of Mathematical Sciences\\
University of Technology Sydney\\
Ultimo NSW, Australia 2007\\
Email: Pavel.Shevchenko@csiro.au
}

\begin{document}
\maketitle

\begin{abstract}
  The Local Volatility model is a well-known extension of the
  Black-Scholes constant volatility model whereby the volatility is
  dependent on both time and the underlying asset. This model can be
  calibrated to provide a perfect fit to a wide range of implied
  volatility surfaces. The model is easy to calibrate and still very
  popular in FX option trading. In this paper we address a question of
  validation of the Local Volatility model. Different stochastic
  models for the underlying can be calibrated to provide a good fit to
  the current market data but should be recalibrated every trading
  date. A good fit to the current market data does not imply that the
  model is appropriate and historical backtesting should be performed
  for validation purposes. We study delta hedging errors under the
  Local Volatility model using historical data from 2005 to 2011 for
  the AUD/USD implied volatility. We performed backtests for a range
  of option maturities and strikes using sticky delta and
  theoretically correct delta hedging. The results show that delta
  hedging errors under the standard Black-Scholes model are no worse
  than that of the Local Volatility model. Moreover, for the case of
  in and at the money options, the hedging error for the Back-Scholes
  model is significantly better.
\end{abstract}


%

\section{Introduction}
Under the well-known Black-Scholes pricing model
\cite{BlackScholes73}, the asset price \( S \) is modelled with
geometric Brownian motion,
\begin{equation}
  \label{eq:1}
  d S_t = \mu_t S_t dt + \sigma S_t dW_t\ ,
\end{equation}
where \( \mu_t \) is the drift, \( \sigma \) is volatility and \( W_t
\) is a standard Brownian motion. One of the key assumptions of the
model is the \emph{no-arbitrage condition}, which means that it is
impossible to make a riskless profit. From this assumption it can be
shown that the fair price of a derivative security with underlying
asset \( S \) is equal to the mathematical expectation of the
discounted payoff of the derivative. This expectation is computed with
respect to a so-called \emph{risk-neutral probability
  measure}. Furthermore, under this measure the dynamics of the asset
price \( S \) is given by
\begin{align}
  \label{eq:2}
  dS_t = (r-q)S_t dt + \sigma S_t dB_t,
\end{align}
where \( B_t \) is a standard Brownian motion under the risk neutral
measure, \( r \) is the constant risk-free interest rate and \( q \)
denotes the constant dividend yield. Given the price of an European
option, strike, maturity and interest rates, the asset price volatility
\( \sigma \) can be computed numerically from the Black-Scholes
pricing formula. We say that \( \sigma \) is the volatility
\emph{implied} by the market price.  If the Black-Scholes model was a
perfect representation of the market, then the implied volatility
would be equal for all market traded options. This is definitely not
the case in practice.

The implied volatility is heavily dependent on the strike price and
maturity of the option. The \emph{local volatility} model is an
extension of the Black-Scholes framework which can account for this
dependence, and it does so by making volatility a function of the
current time and current spot price i.e. \( \sigma(S_t,t)
\). \eqref{eq:2} is then replaced by
\begin{equation}
  \label{eq:3}
 d S_t = (r-q) S_t dt + \sigma(S_t,t) S_t dW_t.
\end{equation}

\section{The Black-Scholes setup}
\label{sec:black-scholes-setup}

Let \( V(S,t) \) denote the \emph{discounted} price of a contingent
claim at time \( t \) with underlying asset price \( S(t) \). Let \( r
\) denote the risk-free interest rate, \( q \) denote the dividend
yield and \( \sigma \) the asset price volatility. Within the
Black-Scholes framework, \( V \) satisfies the fundamental partial
differential equation (PDE)
\begin{align}
  \frac{\partial V}{\partial t} + \frac{1}{2}\sigma^2 S^2
  \frac{\partial^2 V}{\partial S^2} + (r-q)S \frac{\partial
    V}{\partial S} - rV = 0, \label{eq:4}
\end{align}
with known boundary condition \( V(S(T),T) \) for the case of European
options. This PDE can be obtained by applying a trading strategy
called \emph{delta hedging}.

For ease of numerical implementation we transform
the above PDE with \( X = \log S \). Routine calculations show
that the transformed PDE is
\begin{align}
  \frac{\partial V}{\partial t} +(r-q - \frac{1}{2}\sigma^2) \frac{\partial V}{\partial X} +
  \frac{\sigma^2}{2}  \frac{\partial^2 V}{\partial X^2} - rV = 0. \label{eq:10}
\end{align}

\subsection{The Black-Scholes pricing formula}
\label{sec:black-schol-pric}
Assuming constant interest rates \( r_c \), dividend yields \( q_c \)
and volatility \( \sigma_c \), the Black-Scholes formula for the price
of an European call option with strike \( K \) and maturity \( T \),
at zero time \( t=0 \), is given by
\begin{align}
\label{eq:11}
  C&(\sigma_c, K,T,r_c,q_c, S_0)\notag\\
  &= e^{-r_c T} \left(S_0 e^{(r_c-q_c)T} \Phi(d_1) - K
  \Phi(d_1 - \sigma_c\sqrt{T})\right),
\end{align}
where \( \Phi(\cdot) \) denotes the cumulative distribution function
for standard Normal distribution and 
\begin{align*}
  d_1 =
\frac{\ln \left(\frac{S_0}{K}\right) + \left( r_c-q_c + \frac{1}{2}
    \sigma_c^2 \right) T}{\sigma_c \sqrt{T}} .
\end{align*}

For the case where interest rates, dividend yields and volatility are
time-dependent, the Black-Scholes formula can be applied with the
following substitutions (\cite[sec. 8.8]{wilmott2007paul} )
\begin{align}
  r_c &= \frac{1}{T} \int_0^T r(u) du, \label{eq:12}\\
  q_c &= \frac{1}{T} \int_0^T q(u) du, \label{eq:13}\\
  \sigma^2_c &= \frac{1}{T} \int_0^T \sigma^2(u) du,\label{eq:14}
\end{align}
where \( r(t) \), \( q(t) \) and \( \sigma(t) \) are called the
instantaneous interest rate, instantaneous dividend yield and
instantaneous volatility respectively. The interpretation is that in a
small interval of time \( [t, t+\Delta t] \), the amount of interest
accrued/owed is \( r(t) \Delta t \). Note that \( r(t) \), \( q(t) \)
and \( \sigma(t) \) are not observable in the market but instead we
observe the integrals in \eqref{eq:12}, \eqref{eq:13} and
\eqref{eq:14}; see sections \ref{sec:interest-rates} and
\ref{sec:term-struct-volat} for more discussion on this point and for
how to input the correct values from market data.

\section{Local volatility}
\label{sec:local-volatility}

The local volatility model extends the Black-Scholes framework by
making volatility a function of current asset price and time. In
addition we introduce time dependence for the interest rate and
dividend yield. This leads to the following modification of equation
\eqref{eq:10}, 
\begin{align}
  \frac{\partial V}{\partial t} &+(r(t)-q(t) -
  \frac{1}{2}[\sigma(e^{X_t}, t)]^2) \frac{\partial V}{\partial X}
  \notag \\
  &+ \frac{1}{2}[\sigma(e^{X_t}, t)]^2  \frac{\partial^2 V}{\partial X^2} - r(t) V = 0, \label{eq:15}
\end{align}
where the \emph{local volatility} \( \sigma(K,T) \) is given by
\begin{align}
  \sigma&(K,T) \notag \\
&= \sqrt{\frac{2\theta T \frac{\partial \theta}{\partial T} + 2(r(T) -
    q(T))K \theta T \frac{\partial \theta}{\partial K}  }
    {\left[1 + d_1 K \sqrt{T} \frac{\partial \theta}{\partial K} \right]^2
      + K^2 \theta T \left[\frac{\partial^2 \theta}{\partial K^2} -
        d_1 \left(\frac{\partial \theta}{\partial K}\right)^2 \sqrt{T}
      \right]}}, \label{eq:16}\\
  d_1 &= \frac{\ln \left(\frac{S_0}{K}\right) + \int_0^T [r(t) - q(t)]dt +
    \frac{1}{2} \theta^2 T}{\theta \sqrt{T}}. \label{eq:17}
\end{align}

Here we define \( \theta(K,T) = C^{-1}(V) \), where \( V =
C(\theta, \cdot) \) is given by the Black-Scholes formula
\eqref{eq:11}. That is, \( \theta \) is the market \emph{implied
  volatility} for a vanilla option with strike \( K \) and maturity \(
T \). Equation \eqref{eq:16} is sometimes called the \emph{Dupire
  formula}. For a proof of the formula, see
\cite[p.~49]{Shevchenko01}.

The functions \( r(t) \) and \( q(t) \) are the instantaneous rates;
see section \ref{sec:interest-rates} on how to determine these
functions from market data.

To compute local volatility we require an implied volatility surface
that can be interpolated from market data. There is no universal way
to perform this interpolation. We now describe a simple method that
yields good results for FX data.

\subsection{Interpolating market implied volatility}
\label{sec:interp-mark-impl}

To compute the local volatility function \eqref{eq:16}, we need
partial derivatives of the implied volatility surface \( \theta(K,T)
\). In practice we only have a finite number of market data points,
typically \( 5 \) values for a given maturity and about \( 10 \)
maturities; see table \ref{tab:impliedVols}. We need some interpolating procedure for \(
\theta \). This is an ill-posed problem, and there are a number of
ways to interpolate these data points. We use natural cubic splines to
interpolate across strikes and maturities.

\begin{table}[htbp]
  \centering
    \begin{tabular}{rrrrrr}
    \hline
    \multicolumn{1}{c}{\textbf{Maturity / \( \Delta \)}} &
    \multicolumn{1}{c}{\(\boldsymbol{10\Delta}\)\textbf{Put}} & \multicolumn{1}{c}{\(\boldsymbol{25\Delta}\)\textbf{Put}} & \multicolumn{1}{c}{\textbf{ATM}} & \multicolumn{1}{c}{\(\boldsymbol{25\Delta}\)\textbf{Call}} & \multicolumn{1}{c}{\(\boldsymbol{10\Delta}\)\textbf{Call}} \\\hline
    \multicolumn{1}{c}{\textbf{1 week}} & \multicolumn{1}{c}{9.963\%} & \multicolumn{1}{c}{9.088\%} & \multicolumn{1}{c}{8.450\%} & \multicolumn{1}{c}{8.213\%} & \multicolumn{1}{c}{8.338\%} \\
    \multicolumn{1}{c}{\textbf{1 month}} & \multicolumn{1}{c}{10.913\%} & \multicolumn{1}{c}{10.038\%} & \multicolumn{1}{c}{9.400\%} & \multicolumn{1}{c}{9.163\%} & \multicolumn{1}{c}{9.288\%} \\
    \multicolumn{1}{c}{\textbf{2 months}} & \multicolumn{1}{c}{11.363\%} & \multicolumn{1}{c}{10.488\%} & \multicolumn{1}{c}{9.850\%} & \multicolumn{1}{c}{9.613\%} & \multicolumn{1}{c}{9.738\%} \\
    \multicolumn{1}{c}{\textbf{3 months}} & \multicolumn{1}{c}{11.713\%} & \multicolumn{1}{c}{10.838\%} & \multicolumn{1}{c}{10.200\%} & \multicolumn{1}{c}{9.963\%} & \multicolumn{1}{c}{10.138\%} \\
    \multicolumn{1}{c}{\textbf{6 months}} & \multicolumn{1}{c}{12.155\%} & \multicolumn{1}{c}{11.280\%} & \multicolumn{1}{c}{10.630\%} & \multicolumn{1}{c}{10.430\%} & \multicolumn{1}{c}{10.605\%} \\
    \multicolumn{1}{c}{\textbf{1 year}} & \multicolumn{1}{c}{12.400\%} & \multicolumn{1}{c}{11.525\%} & \multicolumn{1}{c}{10.850\%} & \multicolumn{1}{c}{10.675\%} & \multicolumn{1}{c}{10.850\%} \\
    \multicolumn{1}{c}{\textbf{2 years}} & \multicolumn{1}{c}{12.157\%} & \multicolumn{1}{c}{11.350\%} & \multicolumn{1}{c}{10.750\%} & \multicolumn{1}{c}{10.650\%} & \multicolumn{1}{c}{10.844\%} \\
    \multicolumn{1}{c}{\textbf{3 years}} & \multicolumn{1}{c}{12.013\%} & \multicolumn{1}{c}{11.250\%} & \multicolumn{1}{c}{10.700\%} & \multicolumn{1}{c}{10.650\%} & \multicolumn{1}{c}{10.888\%} \\
    \multicolumn{1}{c}{\textbf{4 years}} & \multicolumn{1}{c}{11.966\%} & \multicolumn{1}{c}{11.225\%} & \multicolumn{1}{c}{10.700\%} & \multicolumn{1}{c}{10.675\%} & \multicolumn{1}{c}{10.935\%} \\
    \multicolumn{1}{c}{\textbf{5 years}} & \multicolumn{1}{c}{11.819\%} & \multicolumn{1}{c}{11.100\%} & \multicolumn{1}{c}{10.600\%} & \multicolumn{1}{c}{10.600\%} & \multicolumn{1}{c}{10.881\%} \\
\hline
    \end{tabular}%
  \caption{An example of AUD/USD market implied volatilities on 12
    April 2005. The spot price for that day was \( S_0 = 0.7735 \).}
  \label{tab:impliedVols}%
\end{table}%

\subsection{Our method to compute local volatility}
\label{sec:our-method-compute}
Suppose we have market data for \( N \) different maturities and that
for each maturity \( M \) options are available. Let \( K^{(i)}_j \)
and \( \theta^{(i)}_j \) denote the strike and implied volatility of
the j-th vanilla option with maturity \( T_i \). 
\begin{enumerate}
\item \textbf{Interpolation across strikes:} For each market maturity
  \( T_i \), \( i \in \left\{1,\dots,N \right\} \) fit a natural cubic spline \(
  y_i(k) \) through
\begin{align*}
  \left(K^{(i)}_1, \theta^{(i)}_1\right), \left(K^{(i)}_2,
    \theta^{(i)}_2\right), \dots, \left(K^{(i)}_M, \theta^{(i)}_M\right). 
\end{align*}
Note that \( y_i'(K) = \frac{\partial \theta}{\partial K}_{| (K,T_i)} \)
and \( y_i''(K) = \frac{\partial^2 \theta}{\partial K^2}_{| (K,T_i)}
\).
\item \textbf{Interpolation across maturities:} To find
  \(\frac{\partial \theta}{\partial K} \) at any given \( (K,T) \) fit
  another natural cubic spline \( z(t) \) through
  \begin{align*}
    \left(T_1, y_1'(K)\right), \left(T_2, y_2'(K)\right), \dots, \left(T_N, y_N'(K)\right). 
  \end{align*}
Then \(  \frac{\partial \theta}{\partial K} = z(T)\).
\item Similarly to find \(\frac{\partial^2 \theta}{\partial K^2} \) at any given \(
  (K,T) \) fit another natural cubic spline \( w(t) \) through 
  \begin{align*}
    \left(T_1, y_1''(K) \right), \left(T_2, y_2''(K) \right), \dots,
    \left(T_N, y_N''(K) \right).
  \end{align*}
Then \(  \frac{\partial^2 \theta}{\partial K^2} = w(T)\).
\item To find \( \theta \) and \(\frac{\partial \theta}{\partial T} \) at \( (K,T) \),
  fit a natural cubic spline \( u(t) \) through
\begin{align*}
    \left(T_1, y_1(K) \right), \left(T_2, y_2(K) \right), \dots, \left(T_N, y_N(K) \right). 
  \end{align*}
Then \( \theta(K,T) = u(T) \) and  \(\frac{\partial \theta}{\partial T} = u'(T)\).
\item Substitute above into \eqref{eq:16} and compute \(
  \left[\sigma(K,T)\right]^2 \). If \( \left[\sigma(K,T)\right]^2 <
  0  \) then we overwrite 
  \begin{align*}
    \sigma(K,T) = 0.
  \end{align*}
\end{enumerate}

Note that the method can obtain a value for local volatility for any
\( (K,T) \) pair beyond the market range (for \( T \) smaller than the
first market maturity, larger than the last market maturity etc), by
linear extrapolation of the natural cubic splines. For example, if we
have a natural cubic spline \( y(x) \) fitted to data points \( x_1,
x_2, \dots, x_n \) then our extrapolation function \( y^*(x) \) is
defined by \( y(x^*) \) for \( x* < x_1
\) by 
\begin{align}
 y^*(x) = \begin{cases}
   y'(x_1) (x - x_1) + y(x_1), \textrm{ if \( x < x_1 \),}  \\
   y(x), \textrm{ if \( x_1 \le x \le x_n \),}\\
   y'(x_n) (x - x_n) + y(x_n) , \textrm{ if \( x > x_n \).}
 \end{cases}
\end{align}

\section{Pricing using Crank-Nicolson method}
\label{sec:pricing-using-crank}

Once we have a computable local volatility function, we can use the finite
difference method to solve the PDE \eqref{eq:15}. Suppose that we
would like to price an European call option with strike \( K \) and
maturity \( T \) years. We approximate the PDE \eqref{eq:15} with
boundary conditions \( V(S(T), T) = (S(T)-K)^+\).

\subsection{Mesh properties}
\label{sec:mesh-properties}

First we need a mesh of \( (\textrm{price}, \textrm{time}) \)
pairs. Suppose that we have \( N \) different time points and \( M \)
different prices in the mesh. Furthermore, assume the mesh is
rectangular and uniformly-spaced with boundaries of
\begin{itemize}
\item \( 0 \) and \( T \) for the time axis.
\item \( S_0 \exp \left\{-D\right\} \) and \( S_0 \exp
  \left\{D\right\} \) for the price axis, where \( D = \gamma
  \bar{\theta}\sqrt{T} \) and \( \bar{\theta} \) is the average of the
  at the money implied volatilities. We set \( \gamma = 7 \) which we
  determined experimentally as a value that resulted in an overall
  small numerical error for the Crank-Nicolson method. This value
  corresponds to a very small probability for the price to move beyond
  \( S_0 e^D\).
\end{itemize}

In addition, we scale the number of time points by \( T \). Setting \(
N = 500T + 500 \) gives sufficiently good results. Define \( \Delta t
= T/N \). Then the time interval \( [0,T] \) is discretised by
\begin{align*}
  t_0 = 0, t_1 = \Delta t, t_2 = 2\Delta t, \dots, t_N = T .
\end{align*} 

Also define \( \Delta x = 2D/M \). Then the price interval \( [S_0
e^{-D}, S_0 e^{D}] \) is discretised by 
\begin{align*}
  s_i = S_0 e^{-D+ i (\Delta
  x)},
\end{align*}
for \( i = 0, 1, \dots, M \). This definition allows mesh points
to coincide with the spot price. We did this for the convenience of the
backtesting procedure to calculate difference vanillas using the same
mesh.  Next, the PDE \eqref{eq:15} contains partial derivatives with
respect to the logarithm of price. Let \( x_i = \ln s_i \). Then
\begin{align*}
  x_{i+1} -x_i 
&= \ln s_{i+1} - \ln s_i \\
&= \ln \left(\frac{S_0 \exp \left\{-D + (i+1) \Delta x\right\}}{S_0 \exp
    \left\{-D + i \Delta x\right\}}\right) \\
&= \Delta x,
\end{align*}
so the price points are indeed uniformly spaced in terms of log-prices.

\subsection{The finite difference scheme}
\label{sec:finite-diff-scheme}

Let \( i \in \left\{1,\dots, N\right\} \), \( \nu(t,x) = r(t) - q(t) -
\frac{1}{2}\left[\sigma(e^x, t)\right]^2 \) and \( V_{j}^{i} = V(s_j, t_i)
\).  The \emph{Crank-Nicolson} scheme is given by
\begin{align}
  a^{i-1}_j V_j^{i-1} &- b^{i-1}_j  V_{j+1}^{i-1} - c^{i-1}_j
  V_{j-1}^{i-1} \notag \\
  &= d^i_j  V_j^{i}
  + b^i_j  V_{j+1}^i + c^i_j  V_{j-1}^i, \label{eq:28} 
\end{align}
where 
\begin{align*}
  a^i_j &= \frac{r(t_{i})}{2} + \frac{1}{\Delta t} +
  \frac{\sigma^2(t_{i},s_j)}{2 (\Delta x)^2}, \\
  b^i_j &= \frac{\sigma^2(t_{i},s_j)}{4 (\Delta x)^2}
  + \frac{\nu(t_{i},s_j)}{4 \Delta x}, \\
  c^i_j &= \frac{\sigma^2(t_{i},s_j)}{4 (\Delta x)^2}
  - \frac{\nu(t_{i},s_j)}{4 \Delta x}, \\
  d^i_j &= \frac{1}{\Delta t} - \frac{r(t_{i})}{2} -
  \frac{\sigma^2(t_{i},s_j)}{2 (\Delta x)^2}.
\end{align*}

For boundary conditions we use the fact that \( \lim_{S \to 0}
\frac{\partial V}{\partial S}= e_0 \) and \( \lim_{S \to \infty}
\frac{\partial V}{\partial S}= e_\infty  \) where 
\begin{align*}
  e_0 &= \begin{cases}
            0, \textrm{ if \( V \) is price of call option}\\
            -1, \textrm{ if \( V \) is price of put option}
            \end{cases} \\ 
  e_\infty &= \begin{cases}
            1, \textrm{ if \( V \) is price of call option}\\
            0, \textrm{ if \( V \) is price of put option}
            \end{cases}
\end{align*}

This leads to the following equations 
\begin{align}
  V_0^i - V_1^i &= e_0(s_0 - s_1), \label{eq:29} \\ 
  V_M^i - V_{M-1}^i &= e_\infty(s_M - s_{M-1}). \label{eq:30}
\end{align}

To initiate the scheme, we set for all \( j = \left\{0, \dots, M\right\} \)
\begin{align*}
  V_j^N   = 
  \begin{cases}
    \left(s_j - K\right)^+, &\mbox{if we are pricing a call option.}\\
    \left(K-s_j\right)^+, &\mbox{if we are pricing a put option.}
  \end{cases}
\end{align*}

We then repeatedly solve the system until we obtain \( \left(V_1^0,
  V_2^0, \dots, V_M^0\right)^\mathrm{T} \) (for details see for
e.g. \cite{wilmott2007paul}). If \( M \) is an odd integer, the price
of the option is \( V_{(M+1)/2}^0 \). Otherwise we may fit an
interpolating function \( \hat{V}(s) \) through
\begin{align*}
  (s_1,V^0_1), (s_2,V^0_2), \dots, (s_M,V^0_M)
\end{align*}
and the price is then \( \hat{V}(S_0) \). Also note that the delta of
the option is \( \hat{V}'(S_0)  \). We have found that a natural cubic
spline for \( \hat{V} \) gives good results.

\textbf{Remark: } This pricing method is very fast if
the meshpoints of our local volatility function coincide with the
meshpoints in our finite difference scheme. This is how we implemented
our scheme; we first set the mesh points for our finite difference
scheme then pre-compute the local volatility function at these points.




\section{Market data layout}
\label{sec:market-data-layout}

In this paper we work with daily AUD/USD implied volatility data dating
from 2005/03/22 to 2011/07/15. For each trading day, the market data
contains a spot price and for a range of maturities (1 week, 1 month,
2 months, 3 months, 6 months, 1 year, 2 years, 3 years, 4 years and 5
years) there are \begin{itemize}
\item implied volatility for at the money (ATM) options;
\item risk reversals for 10 and 25 delta call, denoted by \(
  RR_{10\Delta Call} \) and \( RR_{25\Delta Call}  \) respectively;
\item butterflys for 10 and 25 delta put, denoted by \(
  Fly_{10\Delta Put} \) and \( Fly_{25\Delta Put}  \) respectively;
\item zero rates (yields) for the domestic and foreign currency.
\end{itemize}

From this data we need to extract the strike prices and implied
volatilities for traded vanilla options. This is done through the
Black-Scholes framework. Taking the Black-Scholes price of a call
option \eqref{eq:11} and differentiating, we obtain the \emph{call
  delta}
\begin{align}
  \label{eq:32}
  \Delta_{call}(S_0,K,T,\gamma_d,\gamma_f,\sigma ) 
  &= \frac{\partial C (S_0,K,T,\gamma_d,\gamma_f,\sigma ) }{\partial
    S} \notag \\
  &= e^{-\gamma_f T}\Phi(d_1),  
\end{align}
where \( \gamma_d = \frac{1}{T}\int_0^T f(t)dt\) and \( \gamma_f =
\frac{1}{T}\int_0^T q(t)dt\) denote the domestic and foreign yields
respectively (this is discussed in detail in section
\ref{sec:interest-rates}). Utilising put-call parity the \emph{put
  delta} is given by
\begin{equation}
  \label{eq:33}
  \Delta_{put}(S_0,K,T,\gamma_d,\gamma_f,\sigma )  
= \Delta_{call}(S_0,K,T,\gamma_d,\gamma_f,\sigma )  - e^{-\gamma_f T}.
\end{equation}

Following a standard notation we use \( 10\Delta Put \) and \(
25\Delta Put \) to denote the volatility \( \sigma \) and strike price
\( K \) that gives a put delta \( \Delta_{put} \) of \( 10\% \) and \(
25\% \) respectively. Similarly, \( 10\Delta Call \) and \( 25\Delta
Call \) denotes the volatility \( \sigma \) and strike price \( K \)
that gives a call delta \( \Delta_{call} \) of \( 10\% \) and \( 25\%
\) respectively.

\subsection{Computing implied volatilities}
\label{sec:comp-impl-volat-1}

Using market definitions we reconstruct the market implied volatilities by using the
following formulas,
\begin{align*}
  \sigma_{10\Delta Put} &= \sigma_{ATM} + Fly_{10\Delta} -
  \frac{1}{2}RR_{10\Delta }, \\
\sigma_{25\Delta Put} &= \sigma_{ATM} + Fly_{25\Delta } -
  \frac{1}{2}RR_{25\Delta }, \\
\sigma_{25\Delta Call} &= \sigma_{ATM} + Fly_{25\Delta } +
  \frac{1}{2}RR_{25\Delta }, \\
\sigma_{10\Delta Call} &= \sigma_{ATM} + Fly_{10\Delta } +
  \frac{1}{2}RR_{10\Delta }. 
\end{align*}

\subsection{Computing strikes}
\label{sec:computing-strikes}
After we determine the implied volatilities, the only parameter yet to
be determined is the strike price. We use the delta formulas of
\eqref{eq:32} and \eqref{eq:33} and an implementation of the inverse
cumulative distribution function \( \Phi^{-1} \) of the standard
Normal distribution to determine the strike. For example, to obtain
the strike price for the \( 10\Delta Put \) option we are looking for
the value of \( K \) satisfying
\begin{align*}
  \delta_{put}(S_0,K,T,\gamma_d,\gamma_f,\sigma_{10\Delta Put}  ) = 0.1,
\end{align*}
which is easy to calculate via the inverse Normal distribution
function.

\subsection{Interest rates}
\label{sec:interest-rates}
On each trading day, we can extract from the market so called
\emph{zero-coupon interest rates} or \emph{zero rates} for a range of
different maturities. To explain the meaning of these rates we give an
example. Suppose we have the following market zero rates for the
Australian dollar:

\vspace{5mm}
\begin{tabular}{lc}
\hline
\textbf{Maturity (years)} & \textbf{Zero rate (\%)} \\
1 & 4.8\\
2 & 4.9 \\
3 & 5 \\
4 & 5.1 \\
\hline
\end{tabular}
\vspace{5mm}

With continuous compounding of interest rates, a one year investment of
\$10 AUD grows to \( 10 \times e^{0.048 \times 1} = 10.49\). Two
year investment of the same amount grows to \( 10 \times e^{0.049
  \times 2} = 11.03 \)

We can now interpolate between these data points to obtain what is
called a \emph{zero curve}. There is no universally accepted way to
perform this interpolation. Suppose the market rates are given by \(
(T_1, \gamma_1), (T_2, \gamma_2), \ldots, (T_n, \gamma_n) \) where \(
T_i \) denotes the \( i^{th} \) maturity and \( \gamma_i \) is its
zero rate. We define our zero curve \( \gamma (t) \) to be a function
such that \( t \gamma(t) \) is piecewise-linear through the points 
\begin{align*}
  (0,\gamma_1), (T_1, \gamma_1), (T_2,\gamma_2), \ldots, (T_n,\gamma_n).
\end{align*}
Now we will need to have the instantaneous interest rates for
various calculations such as the Dupire formula \eqref{eq:16}. That
is, we need to find the function \( r(s) \) such that \( \gamma(t) =
\frac{1}{t}\int_0^t r(s)ds \). Since we assumed that \( t \gamma(t) \)
is piecewise-linear, it implies that \( r(s) \) is piecewise-constant
on the same intervals that \( t\gamma(t) \) is piecewise-linear.

By construction \( \gamma(t) = \gamma_1 \) for \( t \in [0,T_1] \)
which implies that \( r(t) = \gamma(t) \) over \( [0,T_1]  \). Next
let \( t \in (T_i, T_{i+1}] \) and consider 
\begin{align*}
  \gamma(T_{i+1}) T_{i+1} - \gamma(T_i) T_i
&= \gamma_{i+1}T_{i+1} - \gamma_i T_i\\
&= \int_{T_i}^{T_{i+1}} r(s)ds\\
&= (T_{i+1} - T_i)r(t),
\end{align*}
so it follows that 
\begin{align*}
  r(t) = \frac{\gamma_{i+1}T_{i+1} - \gamma_i T_i}{T_{i+1} -
    T_i}, \mbox{for \( t \in (T_i, T_{i+1}] \)}.
\end{align*}

In summary, the instantaneous interest rate is given by 
\begin{equation}
  \label{eq:34}
  r(t) = 
\begin{cases}
  \gamma_1, \mbox{for \( t \in [0,T_1] \)}      \\
\frac{\gamma_{i+1}T_{i+1} - \gamma_i T_i}{T_{i+1} -
    T_i}, \mbox{for \( t \in (T_i, T_{i+1}] \), \( i = 1, \ldots,n-1 \)}
\end{cases}
\end{equation}

\subsection{Term structure of volatility for Black-Scholes}
\label{sec:term-struct-volat}

To best compare performance of Black-Scholes with Local Volatility, we
need to have time dependent volatilities for the Black-Scholes
model. Under this condition, the market implied volatilities allow us
to construct the term structure of volatility. Suppose that at the
money implied volatilities are
\begin{align*}
  (T_1, \sigma_1), (T_2, \sigma_2), \ldots, (T_n, \sigma_n)
\end{align*}
Then similarly to the instantaneous interest rate of the previous
section, we define the instantaneous volatility \( \sigma: [0,T_n] \to
R^+ \) by
\begin{equation}
  \label{eq:35}
  \sigma (t) = 
\begin{cases}
  \sigma_1, \mbox{for \( t \in [0,T_1] \)}      \\
\sqrt{\frac{\sigma^2_{i+1}T_{i+1} - \sigma^2_i T_i}{T_{i+1} -
    T_i}}, \mbox{for \( t \in (T_i, T_{i+1}] \), \( i = 1, \ldots,n-1 \)}
\end{cases}
\end{equation}
From this define \( \sigma_{avg} : [0,T_n] \to R^+ \) by \(
\sigma_{avg}(t) = \sqrt{\frac{1}{t}\int_0^t \sigma^2(s) ds} \). Then it's easy
to check that \( \sigma^2_i = \sigma^2_{avg}(T_i)\) for all \( i=1,
\ldots, n \).


\section{Calibrating the model}
\label{sec:calibrating-model}

By calibrating the model, we mean a verification of our procedures
above by comparing our results with market data. This is done by 
\begin{enumerate}
\item Obtain current market data, and construct the local volatility
  surface as outlined in Section \ref{sec:our-method-compute}.
\item For each market traded option \( V \), 
\begin{enumerate}
\item Obtain \( V \)'s strike price and maturity. Then apply the
  pricing methodology in Section \ref{sec:pricing-using-crank} to
  obtain a price.
\item Using the Black-Scholes formula \eqref{eq:11}, compute the
  implied volatility from the obtained price and compare with the
  market implied volatility of \( V \). Ideally the computed implied
  volatility should be the same as the market volatility but because
  of numerical errors we have a slight difference (see section
  \ref{sec:implementation}). We adjusted our procedures to obtain an
  absolute difference less than \( 0.5 \% \).
\end{enumerate}
\end{enumerate}

Table \ref{tab:avgCalibrationError} shows the average absolute
difference for calibration error for our implementation over each day
of historical AUD/USD foreign exchange data.
\begin{table}[h]
\centering
\begin{tabular}{lccccc}
  \hline
  \textbf{Maturity/ \( \Delta \)}&
  \(\boldsymbol{10\Delta}\)\textbf{Put} &
  \(\boldsymbol{25\Delta}\)\textbf{Put} & \textbf{ATM} &
  \(\boldsymbol{25\Delta}\)\textbf{Call} &
  \(\boldsymbol{10\Delta}\)\textbf{Call} \\\hline
  \textbf{1 week} & 0.005 & 0.005 & 0.005 & 0.005 & 0.005 \\
  \textbf{1 month} & 0.005 & 0.005 & 0.005 & 0.005 & 0.005 \\
  \textbf{2 months} & 0.005 & 0.005 & 0.005 & 0.005 & 0.005 \\
  \textbf{3 months} & 0.005 & 0.005 & 0.005 & 0.005 & 0.005 \\
  \textbf{6 months} & 0.005 & 0.005 & 0.005 & 0.005 & 0.005 \\
  \textbf{1 year} & 0.005 & 0.005 & 0.005 & 0.005 & 0.005 \\
  \textbf{2 years} & 0.005 & 0.005 & 0.005 & 0.005 & 0.005 \\
  \textbf{3 years} & 0.005 & 0.005 & 0.005 & 0.005 & 0.005 \\
  \textbf{4 years} & 0.005 & 0.005 & 0.005 & 0.005 & 0.005 \\
  \textbf{5 years} & 0.005 & 0.005 & 0.005 & 0.005 & 0.006 \\
  \hline
\end{tabular}
\caption{Average of absolute calibration errors from historical data (\%)}
\label{tab:avgCalibrationError}
\end{table}


\section{Implementation}
\label{sec:implementation}

All our implementations were written in C++. Numerical errors from our
implementation come from using a finite number of mesh points in the
finite difference method (section \ref{sec:mesh-properties}) as well
as the having truncated boundaries for the mesh.

Note that the pricing method for the local volatility model requires
solving tridiagonal systems of equations for finding the natural cubic
spline and for the finite difference method. There exists an algorithm
to solve the system in linear time, see \cite[sec. 2.4]{NumRep}.

\section{Delta Hedging}
\label{sec:delta-hedging}
Let \( V \) denote the price of an option. The \emph{delta} of the
option is defined as \( \Delta = \frac{\partial V}{\partial S}
\). Delta is a measure of the sensitivity of the option price to
changes in the value of the underlying asset. Under the Black-Scholes
framework, \( \Delta \) can be computed explicitly.

Suppose we have a portfolio of options with stocks as
underlying. \emph{Delta hedging} is a strategy to reduce the risk of
the portfolio to changes in price of the underlying assets. To hedge a
short position of one call option we need to take a long position of
\( \Delta \) shares of the underlying asset. Because a change in share
price leads to a change in delta, we must \emph{rebalance} our long
position to maintain the hedge. This means that if the current \(
\Delta \) changes to \( \Delta' \), we must buy or sell to be long \(
\Delta' \) shares. Under the Black-Scholes framework, the rebalancing
must be performed continuously in time to obtain a riskless portfolio.

\subsection{The delta hedging procedure}
\label{sec:delta-hedg-proc}

Suppose that we are selling a European call option with expiry at time
\( T \), and that we wish to rebalance at \( N \) evenly-spaced points
in time. Let \( \delta t = T/N \). Let \( t_0 = 0, t_1 = \delta t,
\dots, t_N = T \). For \( i \in \left\{1, \dots,N\right\} \) let \( S_i \)
and \( \Delta_i \) denote the share price and delta of the call option
at time \( t_i \). Also let the price of a call option at time \( t_0
\) be \( C \). We note that calculation of \( C \) and \( \Delta_i \)
depends on the model we use for the asset price. Let \( r(t) \) and \(
q(t) \) denote the instantaneous interest rate and instantaneous
dividend yield, respectively at time \( t \). These rates are computed
by the formulas in section \ref{sec:interest-rates}. At time \( t_0=0
\) we
 \begin{enumerate}
\item Short one call for \( C \) cash, and go long \( \Delta_0 \)
  shares. The cash position at this time is \(P_0= C - \Delta_0 S_0 \).
\item At \( t_1 \) we perform our first rebalancing. At this point in
  time, we need to be long \( \Delta_1 \) shares which results in a
  cashflow of \( \left(\Delta_0 - \Delta_1\right)S_1 \) . To see why,
  suppose that \( \Delta_0 < \Delta_1 \). We need to buy
  \( \left(\Delta_1 - \Delta_0\right) \) shares which has a cash flow of \(
  -\left(\Delta_1 - \Delta_0\right)S_1 = \left(\Delta_0 - \Delta_1\right)S_1
  \). On the other hand if \( \Delta_1 < \Delta_0 \) we need to sell
  \( \Delta_0 - \Delta_1 \) shares which has a cashflow of
  \( \left(\Delta_0 - \Delta_1\right)S_1 \).
\item Next note that interest charged/accrued on the cash position of
  \( C - \Delta_0 S_0 \) between time \( t_0 \) and \( t_1 \) is
  \( \left(e^{r(t_0)\delta t}-1\right) P_0 \). Similarly, the continuous
  dividend yield paid/received over this time period is
  \( \left(e^{q(t_0)\delta t}-1\right)\Delta_0 S_0 \). 
\item After rebalancing at \( t_1 \), our cash position is 
  \begin{align*}
    P_1 =\ 
  &e^{r(t_0)\delta t}P_0 + (e^{q(t_0)\delta t}-1)\Delta_0S_0\\
  &+ \left(\Delta_0 - \Delta_1\right)S_1. 
  \end{align*}
\item At time \( t_2 \) we need to be long \( \Delta_2 \) shares,
  which results in a cashflow of \( (\Delta_1 - \Delta_2)S_2 \). Again
  taking into account interest and dividend yield, our cash position
  at this time is 
  \begin{align*}
    P_2 =\ &e^{r(t_1) \delta t}P_1 + \left(e^{q(t_1)\delta t}-1\right)\Delta_1S_1\\
  &+ \left(\Delta_1 - \Delta_2\right)S_2.
  \end{align*}
\item In general, the cash position at time \( t_i \), \( i \in
  \left\{1,\dots, N-1\right\} \) is 
  \begin{align*}
    P_i =\ &e^{r(t_{i-1}) \delta t}P_{i-1} + \left(e^{q(t_{i-1})\delta t}-1\right)\Delta_{i-1}S_{i-1}\\
  &+ \left(\Delta_{i-1} - \Delta_i\right)S_i.
  \end{align*}
\item After rebalancing at time \( t_{N-1} \) we have a cash position of \(
  P_{N-1} \) and a long position of \( \Delta_{N-1} \) shares.
\item At maturity \( t_N = T \), we will sell our long position of
  shares. We still earn/pay interest and dividend over the period \(
  [t_{N-1},t_N] \). The final cash position is then 
  \begin{align*}
    P_N =\ 
  &e^{r(t_{N-1})\delta t}P_{N-1} \\
&+ \left( e^{q(t_{N-1}) \delta t} - 1\right)\Delta_{N-1}S_{N-1} +
  \Delta_{N-1}S_N.
  \end{align*} 
The hedging error is then defined as \( P_N
  - \left(S_T - K\right)^+ \).
\end{enumerate}

\subsection{Simulated delta hedging}
\label{sec:simul-delta-hedg}

Under the Black-Scholes model, the asset price \( S \) follows
geometric Brownian motion where it is possible to have time dependent
drift and volatility \( \mu_t \) and \( \sigma_t \). To simulate \( S
\) we use the scheme (see \cite{Glasserman})
\begin{equation}
  \label{eq:36}
  S(t_{n+1}) = S(t_n) \exp \left\{(\mu-\frac{1}{2} \sigma^2)\delta t + \sigma_{t_n} \sqrt{\delta t }Z_n\right\},
\end{equation}
where \( Z_n\) are independent and identically distributed Normal
random variables with mean \( 0 \) and variance \( 1 \).  Using this
scheme to generate a trajectory of the price process \( S \) we may
then perform delta hedging.

Under the local volatility model we may simulate the asset price
process \( S \) by 
\begin{equation}
  \label{eq:37}
  S(t_{n+1}) = S(t_n) \exp \left\{(\mu-\frac{1}{2} \sigma^2)\delta t + \sigma(S(t_n),t_n) \sqrt{\delta t }Z_n\right\}.
\end{equation}

We performed simulated delta hedging under both the Black-Scholes and
local volatility models and observed that hedging errors converged to
zero as the time step decreases to zero.

\section{Historical delta hedging}
\label{sec:hist-delta-hedg}

In this section we apply the delta hedging procedure with real daily
AUD/USD implied volatility data as described in section
\ref{sec:market-data-layout}. For a given call option with maturity of
\( T \) years we set \( N \) be the number trading days between the
day the option is written and the day of maturity. So just like in
section \ref{sec:delta-hedg-proc}, we define \( \delta t = T/N \) and
\( t_i = i \delta t \) for \( i =0, \ldots, N \). Then \( t_i \)
represents the start of the \( (i+1)^{th} \) trading day.

The instantaneous interest rates \( r(t) \) and \( q(t) \) represent
the domestic (AUD) and foreign (USD) rates respectively. The procedure
for the historical backtest is the same as that described in section
\ref{sec:delta-hedg-proc}. However we must be careful with the
interest rates since for each trading day a new sequence of market
zero rates are quoted. To be precise the quantity \( r(t_i) \) that is
needed in the delta hedging procedure is obtained by taking the
domestic zero rate for the nearest quoted maturity \( T_1 \)
\emph{from the market data corresponding to trading day} \( t_i \)
(note the form of equation \eqref{eq:34}). 

The only points where the Black-Scholes and local volatility methods
differ is the calculation of delta \( \Delta_i \) on each trading day.

\subsection{Backtesting under the Black-Scholes framework}
\label{sec:backt-under-black}
For each trading day \( t_i \), define 
\begin{align*}
  \gamma^{(i)}_d &= \frac{1}{T - i\delta t}\int_0^{T - i\delta t} r(t) dt, \\
  \gamma^{(i)}_f &= \frac{1}{T - i\delta t}\int_0^{T - i\delta t} q(t) dt, \\
  \sigma^{(i)}_{avg} &= \sqrt{\frac{1}{T - i\delta t} \int_0^{T -
      i\delta t} \sigma^2(t) dt},
\end{align*}
where these quantities are obtained from the market data at time \(
t_i \). We note that at the money implied volatilities were used.

At time \( t_0 \) we compute the initial call option price \( C(S_0,
K,T, \gamma^{(0)}_d, \gamma^{(0)}_f, \sigma^{(0)}_{avg}) \) by the
Black-Scholes formula \eqref{eq:11}, and initial delta \(
\Delta_{call}(S_0, K,T, \gamma^{(0)}_d, \gamma^{(0)}_f,
\sigma^{(0)}_{avg}) \) given by equation \eqref{eq:32}. To obtain the
cash position \( P_i \) at time \( t_i \) we need the value of delta
\( \Delta_i \) at this time which is computed as
\begin{align*}
  \Delta_i = \Delta_{call}(S_i, K,T- i \delta t, \gamma^{(i)}_d,
\gamma^{(i)}_f, \sigma^{(i)}_{avg}).
\end{align*}

\subsection{Backtesting under the local volatility framework}
\label{sec:backt-under-local}

Under the LV model, we use the finite difference scheme of section
\ref{sec:finite-diff-scheme} to compute the initial option
price. Recall from section \ref{sec:finite-diff-scheme} that the
finite difference scheme results with a sequence of time \( t_0=0 \)
prices \( \left(V(s_1,t_0), V(s_2,t_0), \dots, V(s_M,t_0)\right)
\), where \( s_i \) denotes the price grid points of the
scheme. Fitting an interpolating function \( \hat{V}(s) \) through
\begin{align*}
  (s_1,V(s_1,t_0)), (s_2,V(s_2,t_0)), \dots, (s_M,V(s_M,t_0)),
\end{align*}
the initial price is then given by \( C=\hat{V}(S_0) \). We also define the time
\( t_0 \) delta of the option by \( \Delta_0=\hat{V}'(S_0) \), the
first derivative of \( V \) at \( S_0 \). Next we will describe two methods
of computing the subsequent deltas \( \Delta_1, \dots, \Delta_{N-1}
\). We first introduce some simplifying notions.

Let \( i = 1, \dots, N-1 \). Suppose we apply the finite differencing
scheme to the market data at time \( t_{-1} \). After iteratively
solving the required system of equations we obtain a sequence of time
\( t_{i-1} \) call option prices
\begin{align*}
  \left(V(s_1,t_{i-1}), V(s_2,t_{i-1}), \dots, V(s_M,t_{i-1})\right).
\end{align*}

We then define the function \( \hat{V}_{(i-1)}(s) \) as the natural
cubic spline passing through
\begin{align*}
  (s_1,V(s_1,t_{i-1})), (s_2,V(s_2,t_{i-1})), \dots, (s_M,V(s_M,t_{i-1})).
\end{align*}

\subsubsection{Theoretically correct delta}
\label{sec:theor-corr-delta}

The so-called \emph{theoretically correct delta} \( \Delta_i \) at
time \( t_i \) for \( i=1, \dots, N-1 \) is defined by \( \Delta_i =
\hat{V}_{(i-1)} '(S_i) \), where \( S_i \) is the spot price at time
\( t_i \).

The idea behind this definition of delta is that if we compute a local
volatility function from current market data with spot price \( S_0
\), a subsequent change in the spot price should not alter the local
volatility function. i.e. 
\begin{align}
  \label{eq:38}
  \sigma(S,t;S_0) = \sigma(S,t;S_0+\Delta S),
\end{align}
for some change in spot price \( \Delta S \). If the local volatility
function fully captured the real diffusion process of the underlying,
then \eqref{eq:38} should prevail. However there are claims in the
literature that this is contrary to common market behaviour, see for
example Hagan \cite{hagan2002managing} and Rebonato \cite{RefWorks:32}. 

\subsubsection{Sticky delta}
\label{sec:sticky-delta}

With \emph{sticky delta}, we assume that a change in the spot price
will not result in a change to the implied volatility and the delta
\cite{RefWorks:33}. That is, the market data implied
volatilities (for e.g. in Table \ref{tab:impliedVols}) which are
expressed in terms of maturity and delta do not change when the spot
price changes. This leaves the strike price to be altered. If \(
\hat{\sigma} \) denotes the implied volatility, it can be showed that
\begin{align*}
  \hat{\sigma}(K,T;S_0) = \hat{\sigma}\left(K + \frac{K\Delta
      S}{S_0},T ; S_0 + \Delta S\right).
\end{align*}

That is, under sticky delta a shift in the spot price \( S_0 \) by \(
\Delta S \) leads to a shifting of the market strike \( K \) by \(
\frac{K\Delta S}{S_0} \). To compute the quantity \( \Delta_i \)
under this assumption at time \( t_i \), first compute the option
price at time \( t_{i-1} \), \( \hat{V}_{(i-1)}(S_i) \). Then take the
market data at time \( t_{i-1} \) and perturb the spot price \(
S_{i-1} \) by a small quantity \( \Delta S = (0.001) S_{i-1} \) (this
quantity cannot be very small nor very large due to large errors
introduced in the calculation of derivative. Our chosen value for \(
\Delta S \) was determined from numerical tests for stability and
accuracy). That is define a new spot price \( S_{i-1}^+ = S_{i-1} +
\Delta S\). Taking \( S_{i-1}^+ \) as the new spot price and without
modifying the implied volatilities, deltas and interest rates
recompute the market strike prices as explained in section
\ref{sec:computing-strikes}. Using this modified market data, compute
a new option price by finite difference and interpolate through the \(
t_{i-1} \) prices with the function \( V^+_{(i-1)}(s) \). Similarly,
define another spot price \( S_{i-1}^- = S_{i-1} - \Delta S\),
recompute a new set of strike prices, compute finite difference and
interpolate through the resulting prices with the function \(
V^-_{(i-1)}(s) \).  The \emph{central difference sticky delta} is then
defined as
\begin{align*}
  \Delta_i = \frac{\hat{V}^+_{(i-1)}(S_i) - \hat{V}^-_{(i-1)}(S_i)}{2\Delta S}.
\end{align*}

%
%

\begin{figure}
  \centering \subfloat[][Histogram]
  {\label{fig:hist_one_week_Put10D}\includegraphics[width=0.52\textwidth]{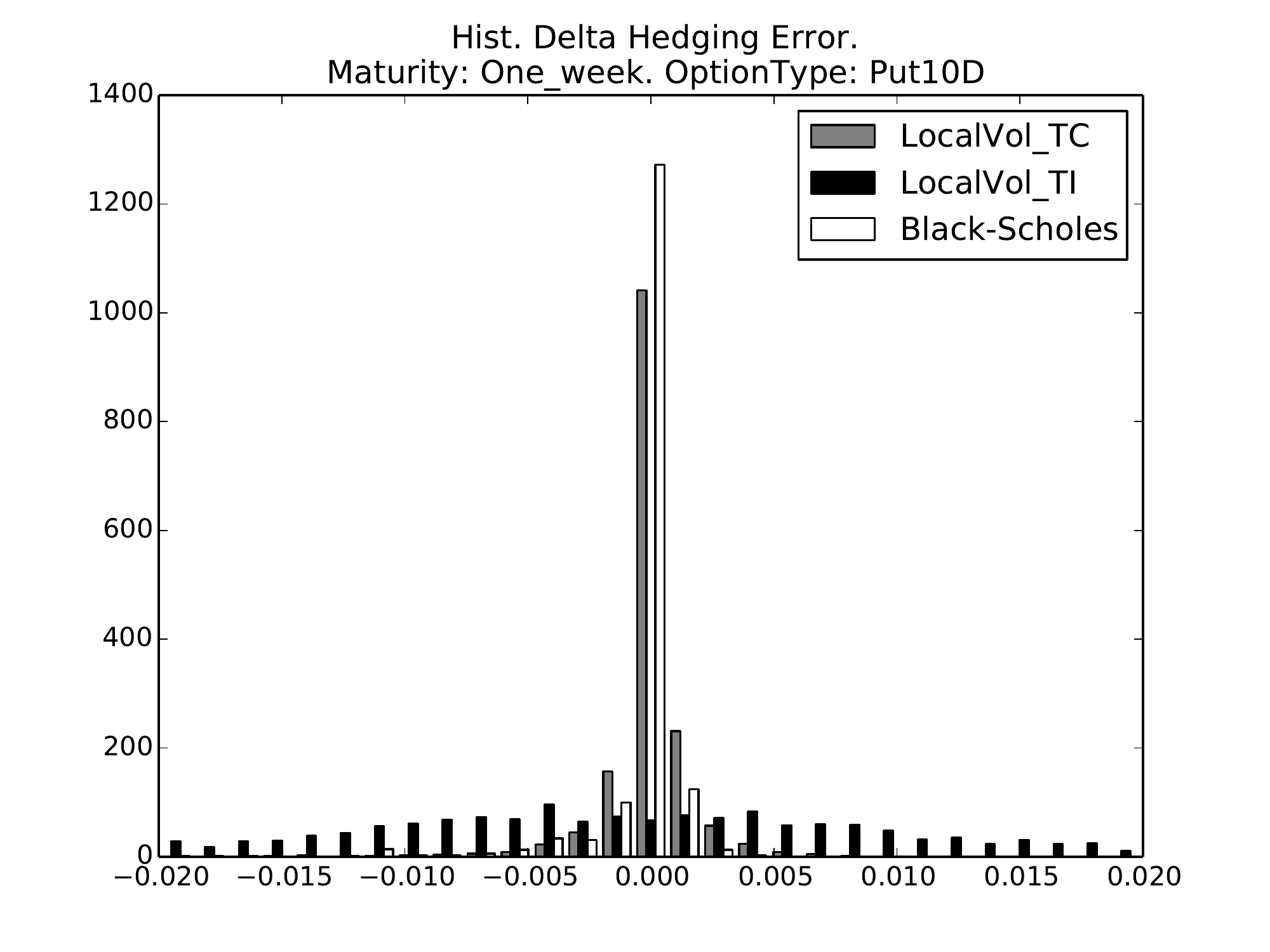}}
  \centering \subfloat[][Hedging error time series]
  {\label{fig:TS_one_week_Put10D}\includegraphics[width=0.52\textwidth]{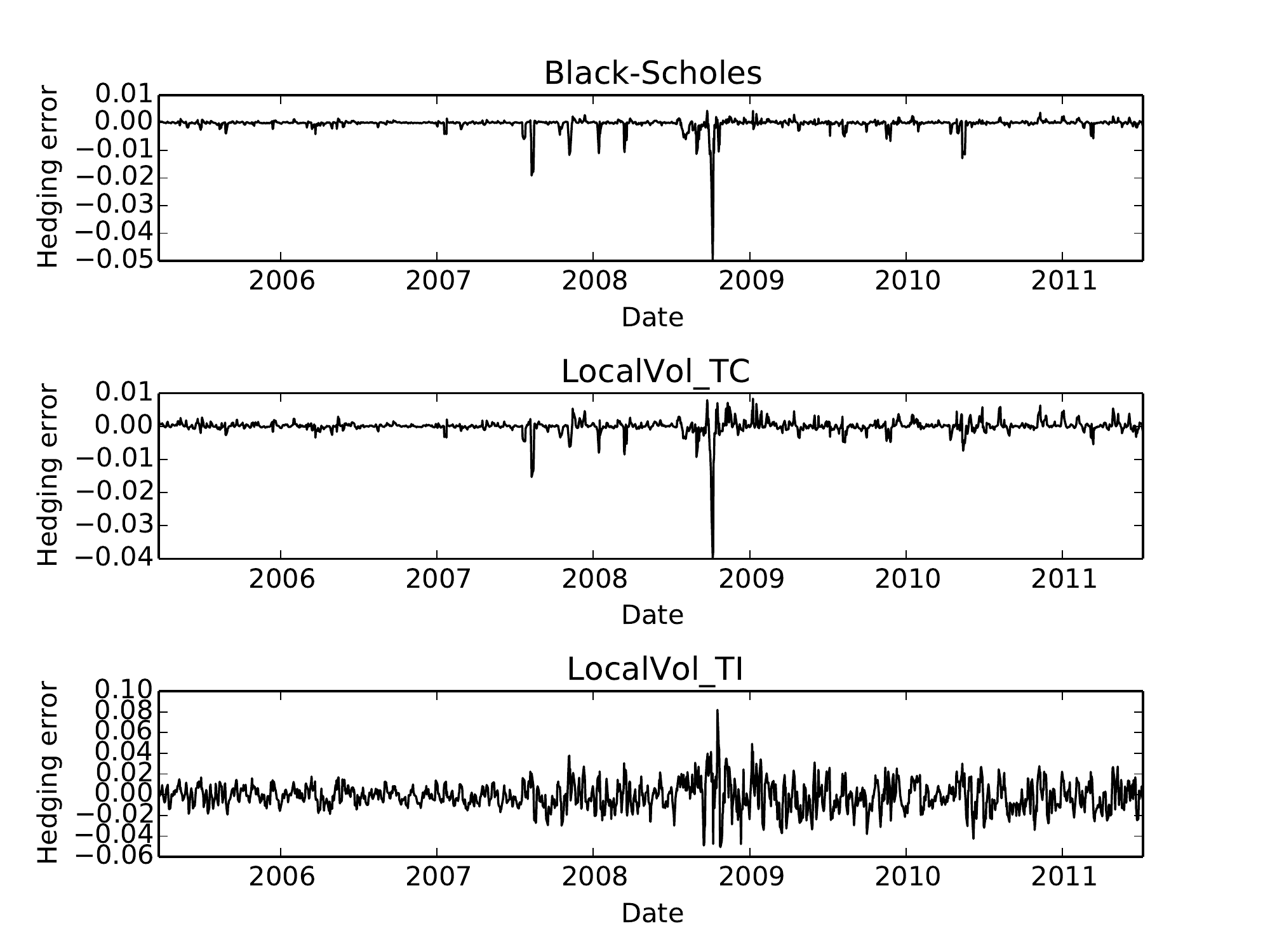}}
  \caption{Delta hedging errors for (10$\Delta$Put) calls with one
    week maturity. LocalVol\_TC and LocalVol\_TI denote
    results from the theoretically correct
    and theoretically incorrect (sticky delta) local volatility models, respectively.}
  \label{figs:hedging_errors_one_week_Put10D}
\end{figure}

\begin{figure}
  \centering \subfloat[][Histogram]
  {\label{fig:hist_one_week_Put25D}\includegraphics[width=0.52\textwidth]{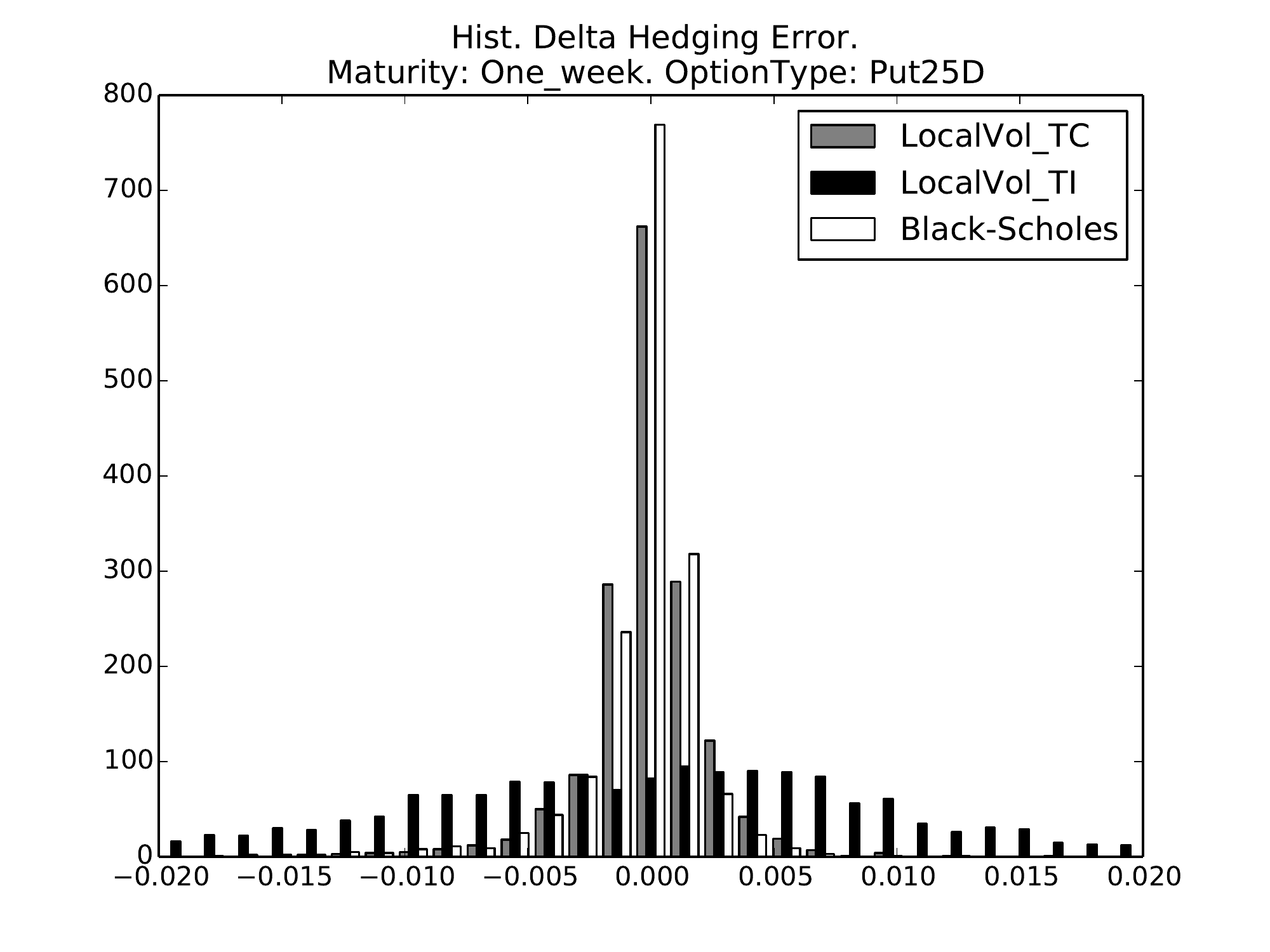}}
  \centering \subfloat[][Hedging error time series]
  {\label{fig:TS_one_week_Put25D}\includegraphics[width=0.52\textwidth]{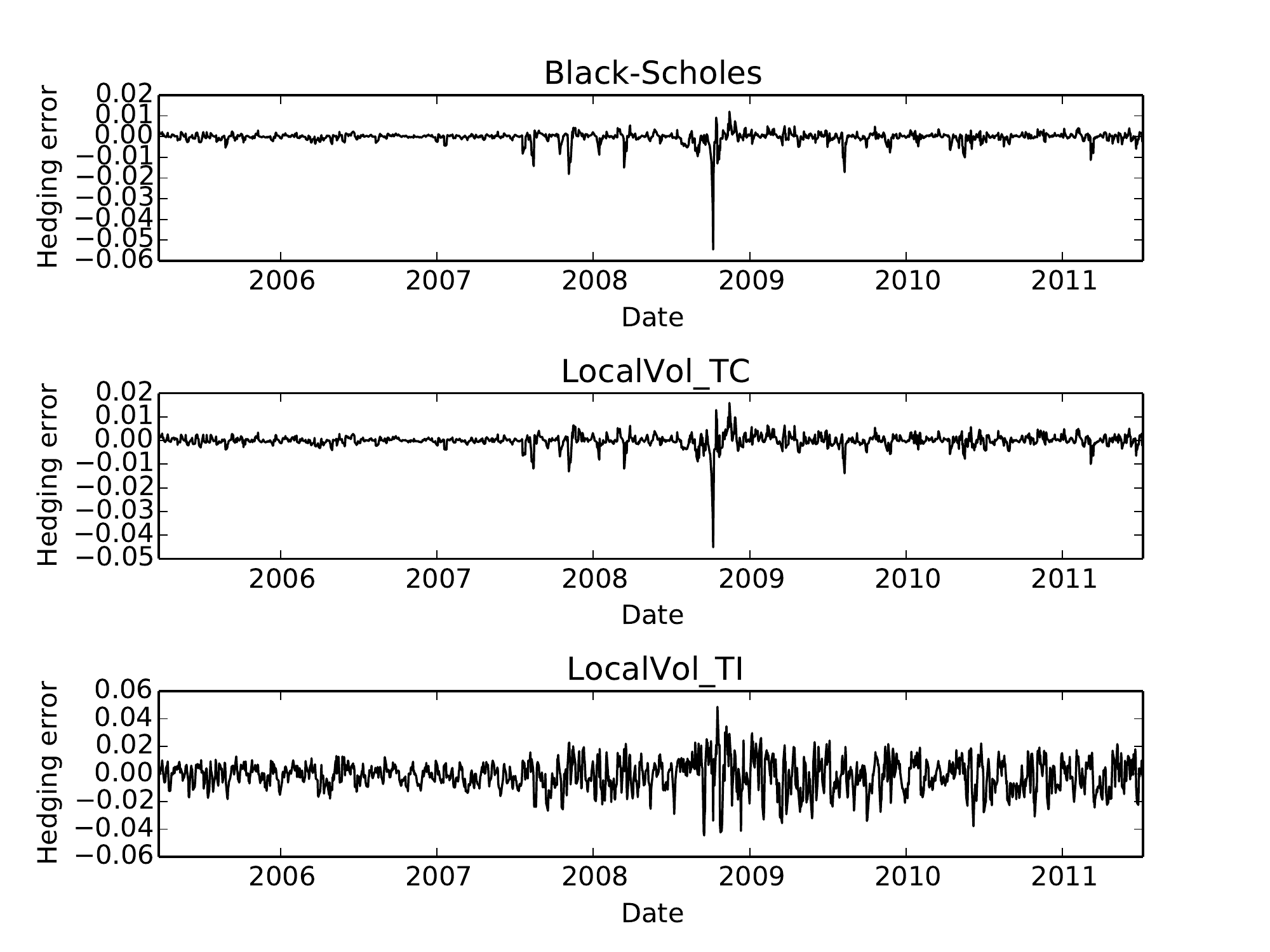}}
  \caption{Delta hedging errors for (25$\Delta$Put) calls with one
    week maturity.}
  \label{figs:hedging_errors_one_week_Put25D}
\end{figure}

\begin{figure}
  \centering \subfloat[][Histogram]
  {\label{fig:hist_one_week_ATM}\includegraphics[width=0.52\textwidth]{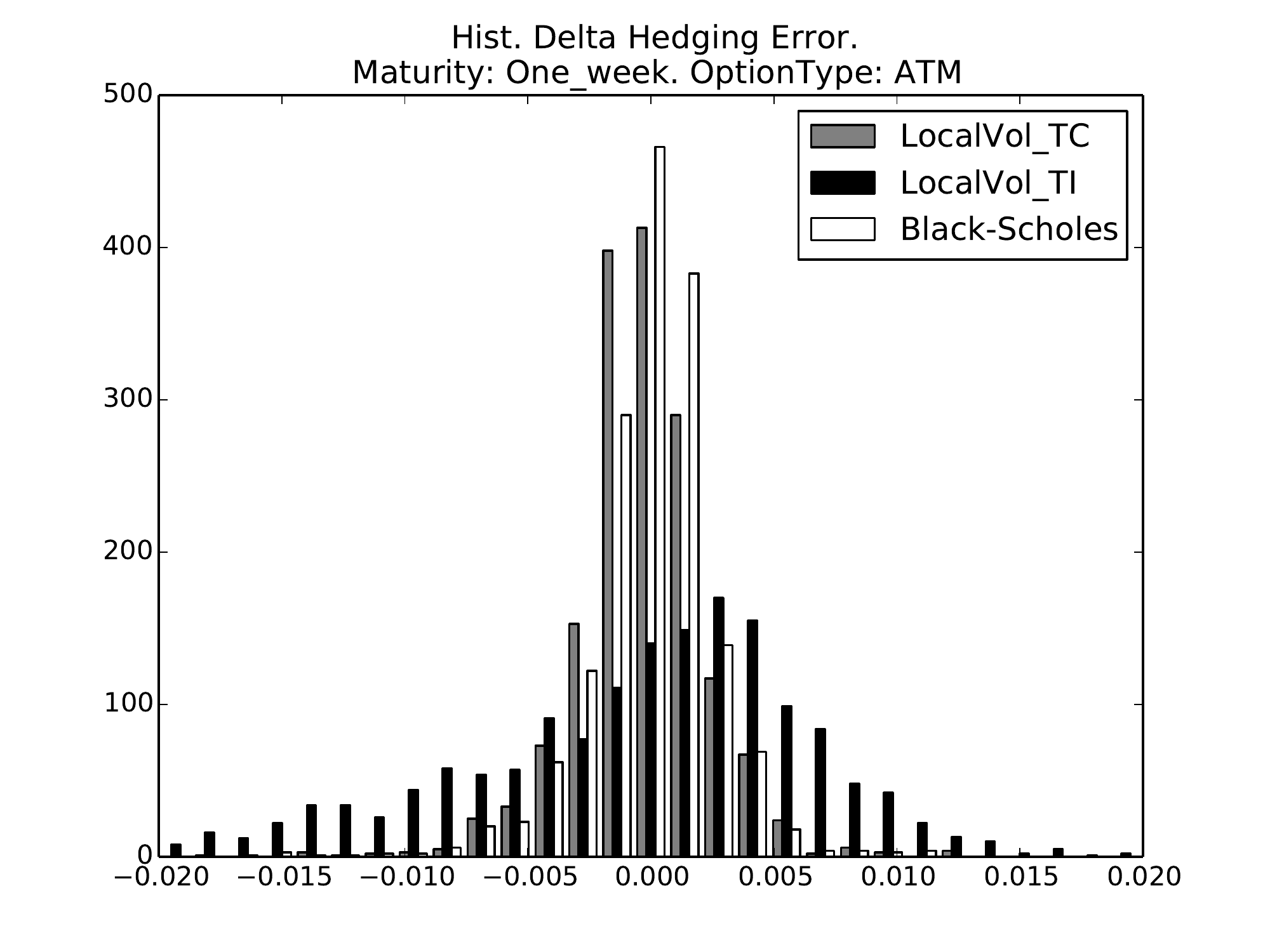}}
  \centering \subfloat[][Hedging error time series]
  {\label{fig:TS_one_week_ATM}\includegraphics[width=0.52\textwidth]{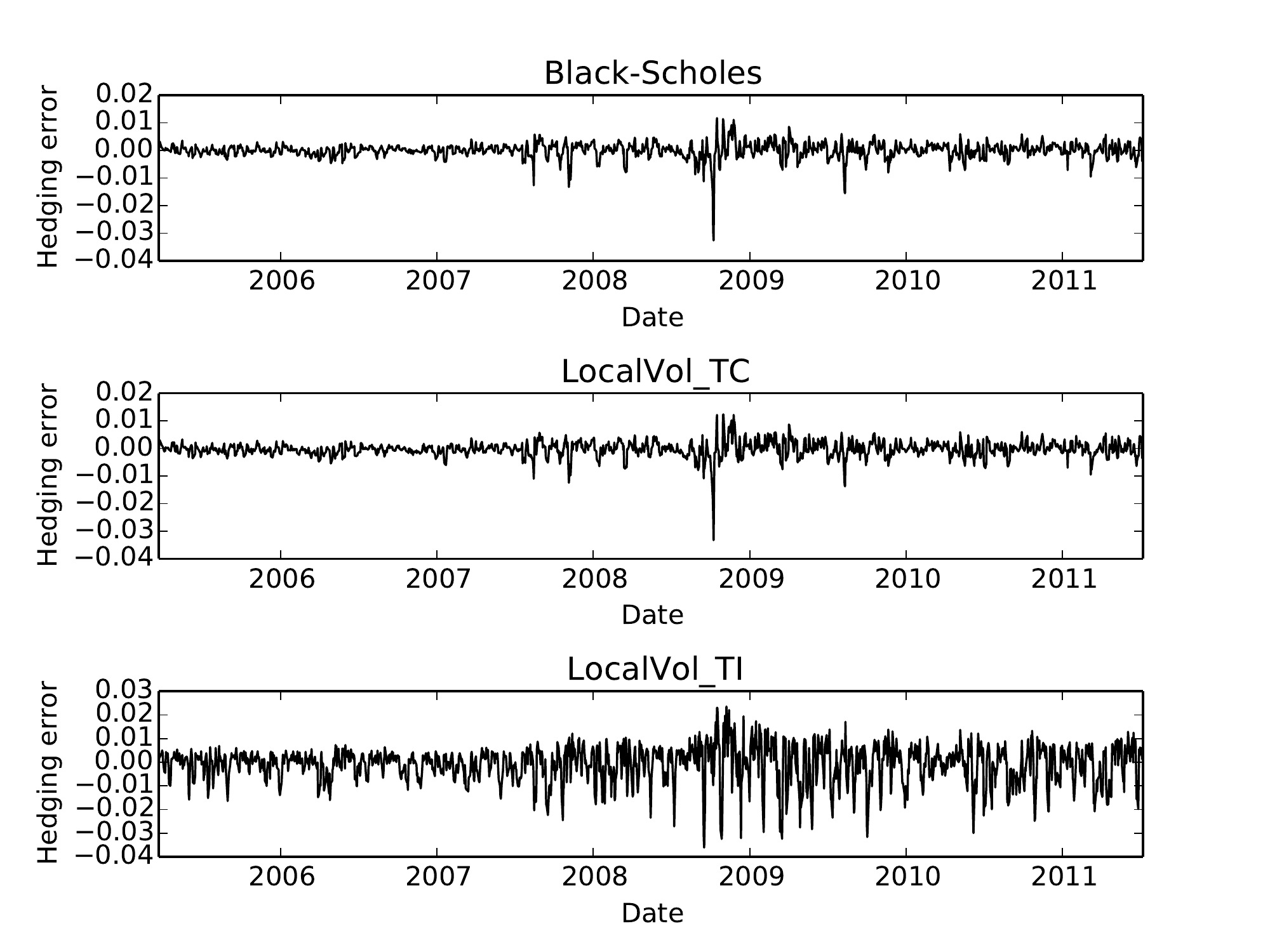}}
  \caption{Delta hedging errors for ATM calls with one
    week maturity.}
  \label{figs:hedging_errors_one_week_ATM}
\end{figure}

\begin{figure}
  \centering \subfloat[][Histogram]
  {\label{fig:hist_one_week_Call25D}\includegraphics[width=0.52\textwidth]{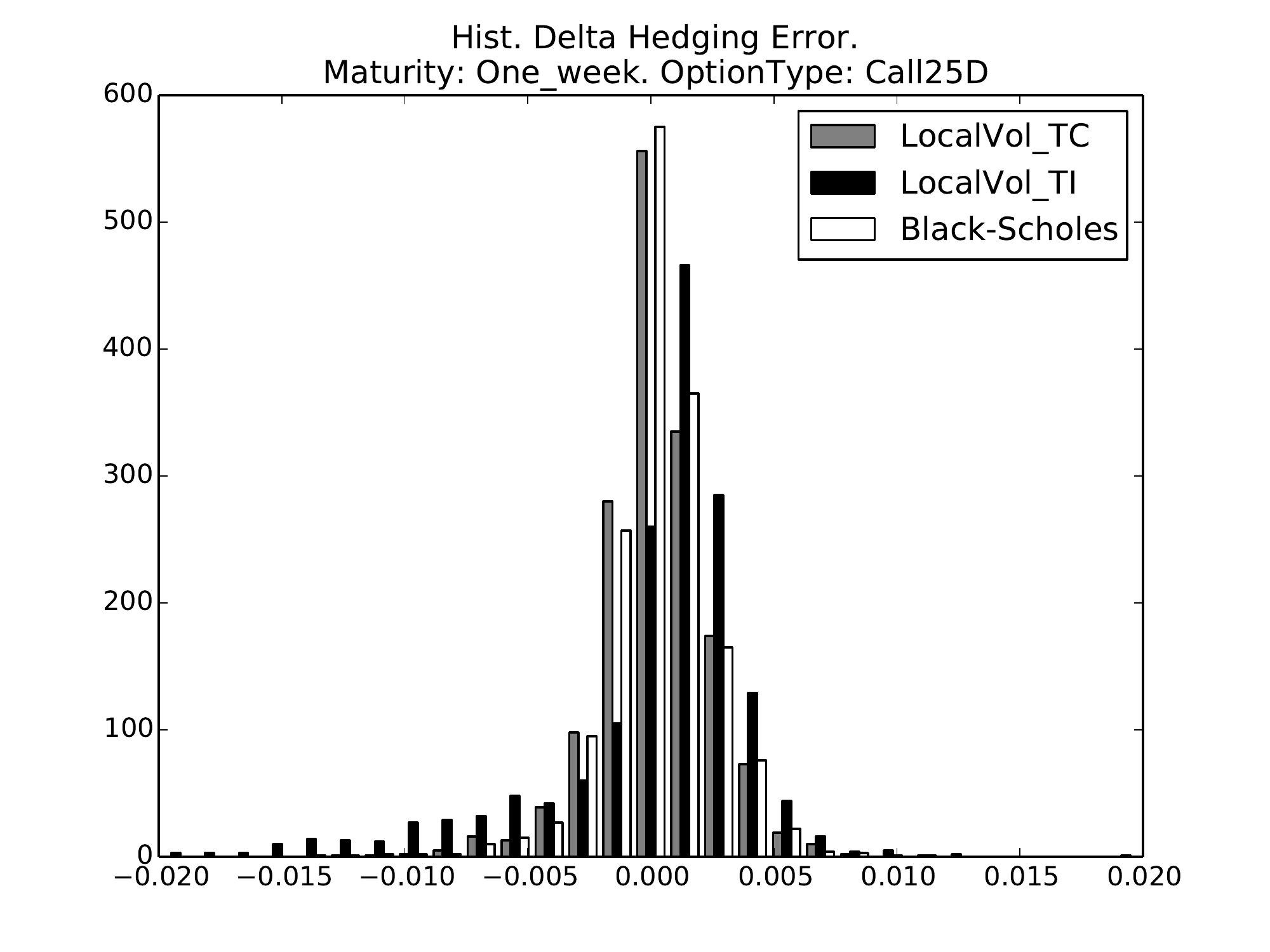}}
  \centering \subfloat[][Hedging error time series]
  {\label{fig:TS_one_week_Call25D}\includegraphics[width=0.52\textwidth]{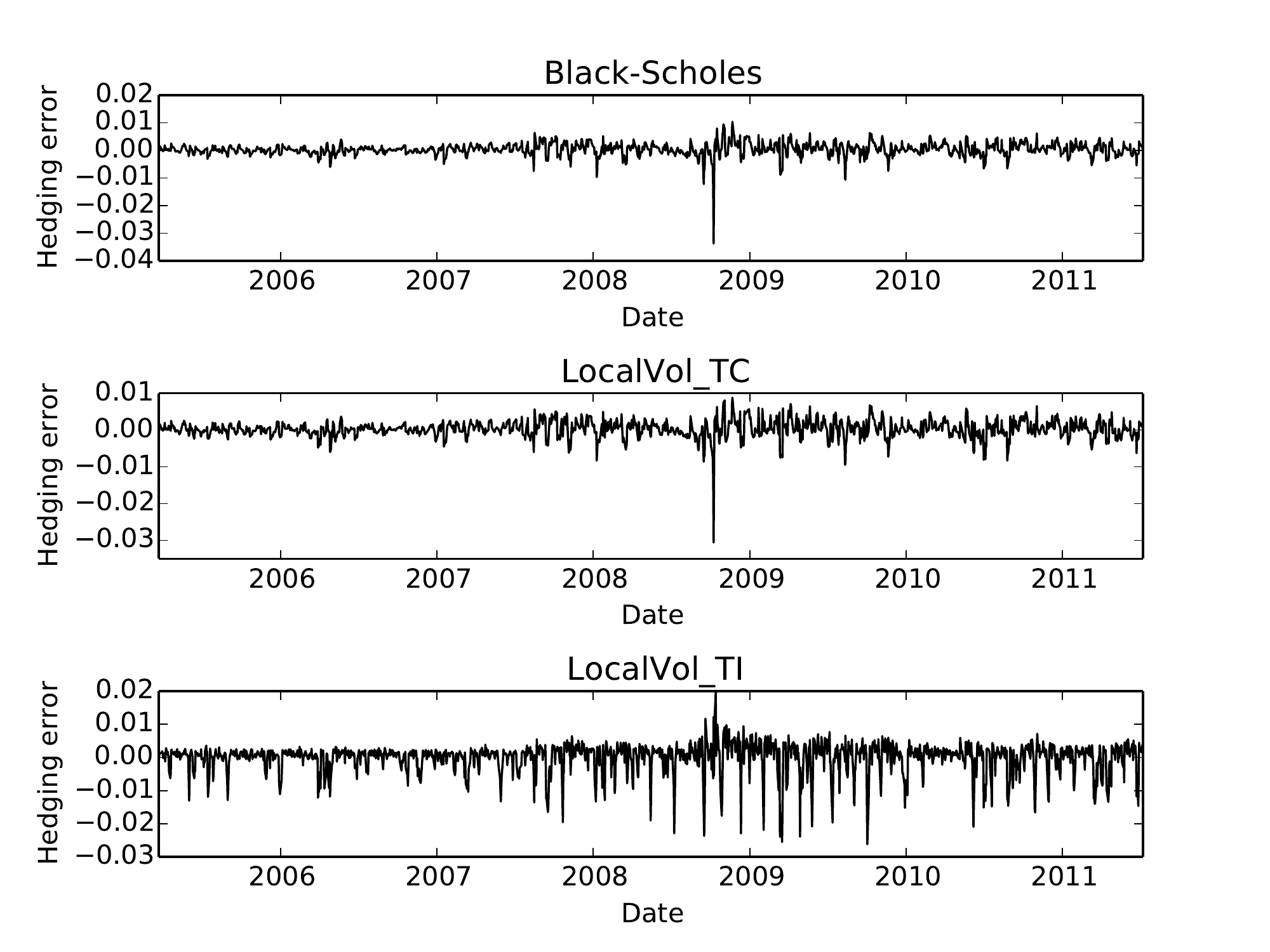}}
  \caption{Delta hedging errors for (25$\Delta$Call) calls with one
    week maturity.}
  \label{figs:hedging_errors_one_week_Call25D}
\end{figure}

\begin{figure}
  \centering \subfloat[][Histogram]
  {\label{fig:hist_one_week_Call10D}\includegraphics[width=0.52\textwidth]{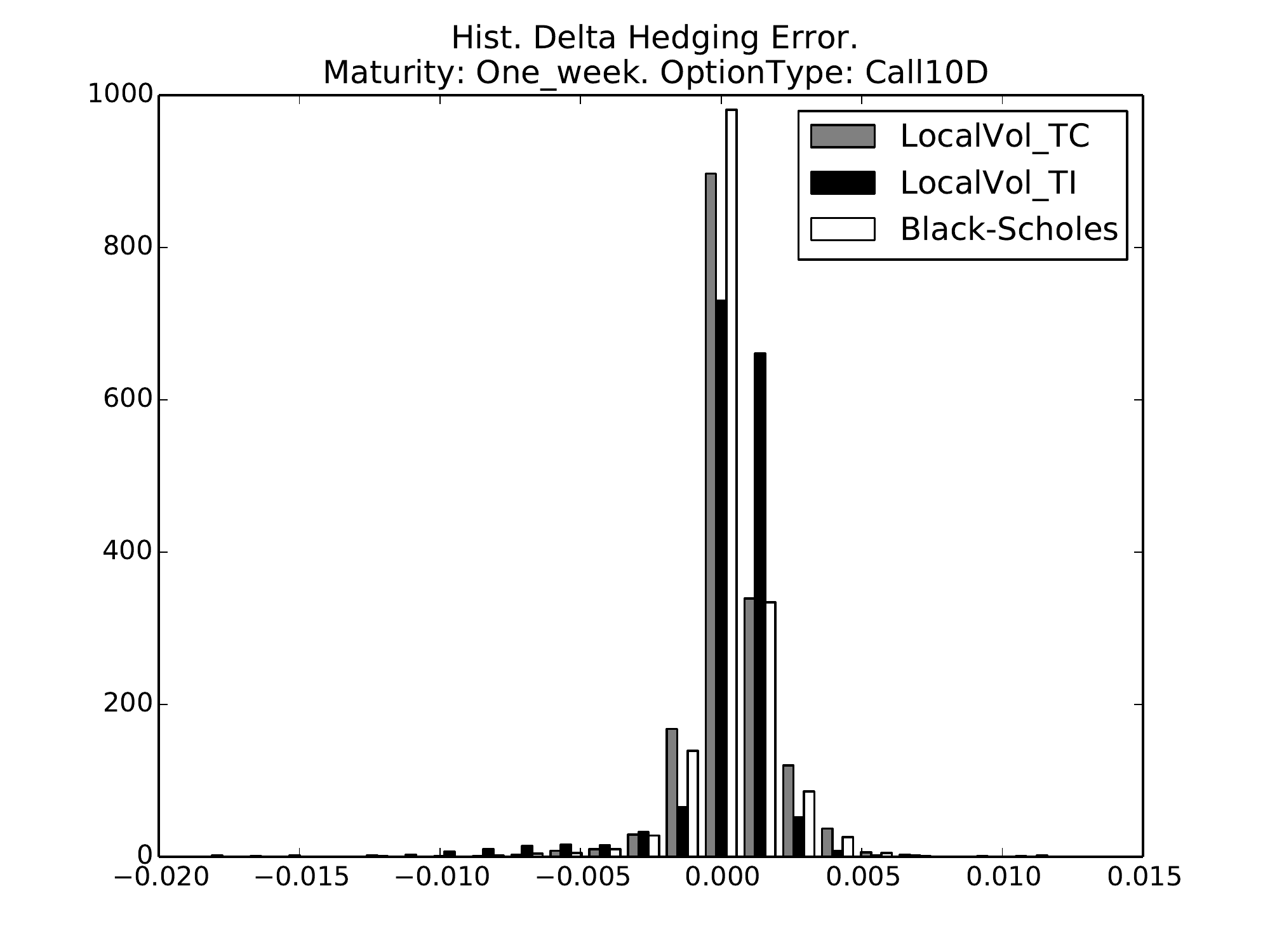}}
  \centering \subfloat[][Hedging error time series]
  {\label{fig:TS_one_week_Call10D}\includegraphics[width=0.52\textwidth]{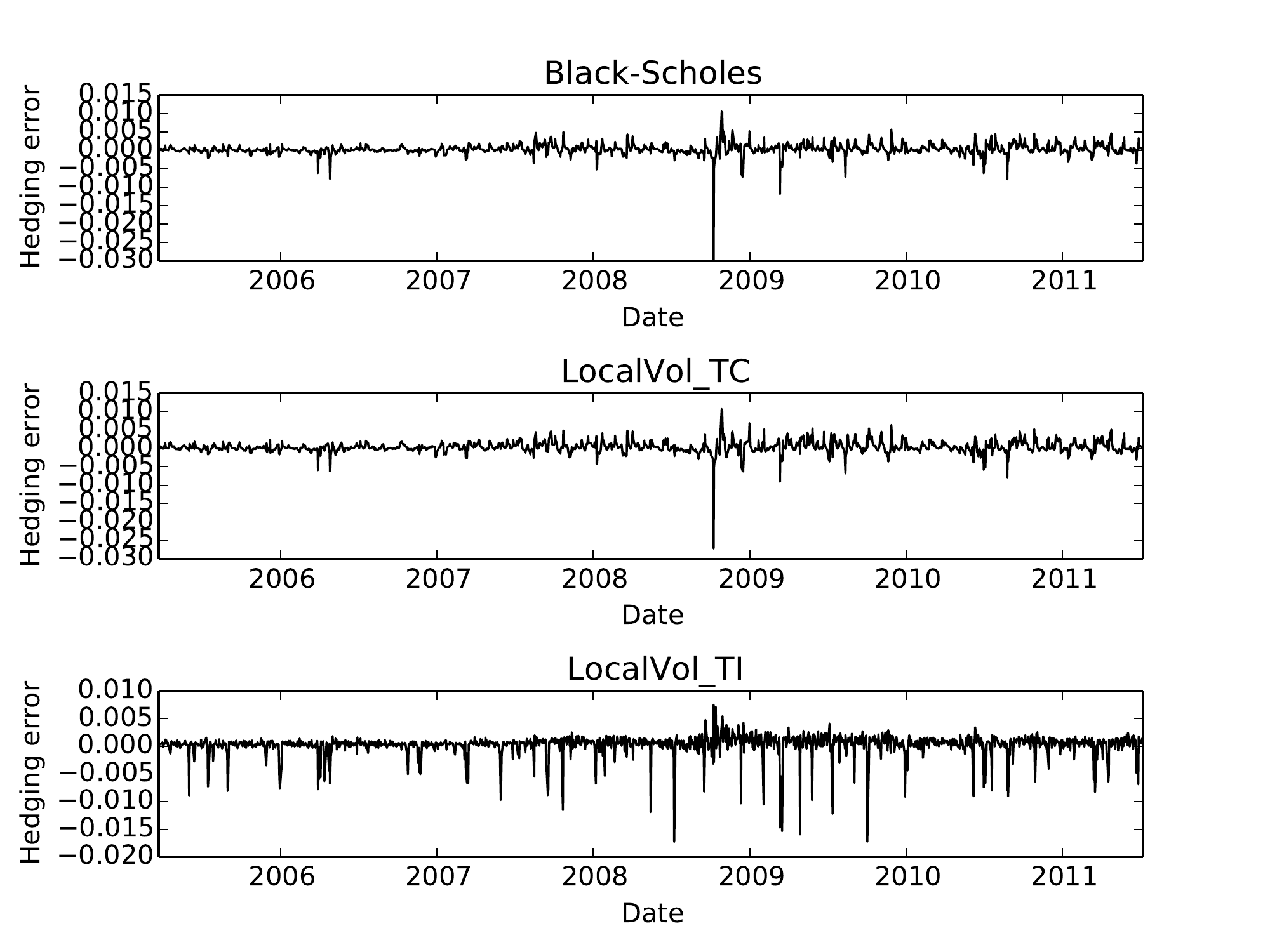}}
  \caption{Delta hedging errors for (10$\Delta$Call) calls with one
    week maturity.}
  \label{figs:hedging_errors_one_week_Call10D}
\end{figure}

\begin{figure}
  \centering \subfloat[][Histogram]
  {\label{fig:hist_one_month_Put10D}\includegraphics[width=0.52\textwidth]{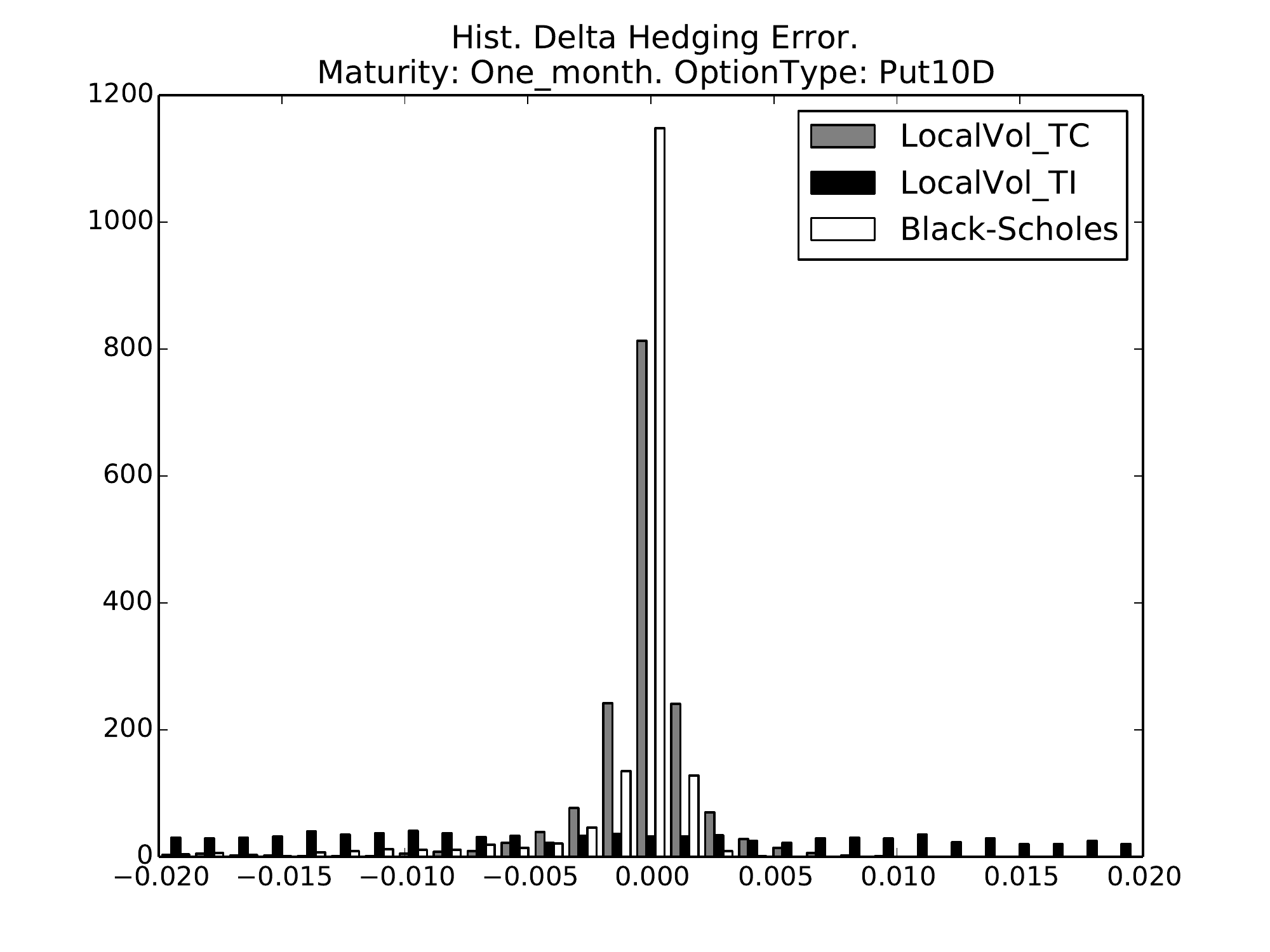}}
  \centering \subfloat[][Hedging error time series]
  {\label{fig:TS_one_month_Put10D}\includegraphics[width=0.52\textwidth]{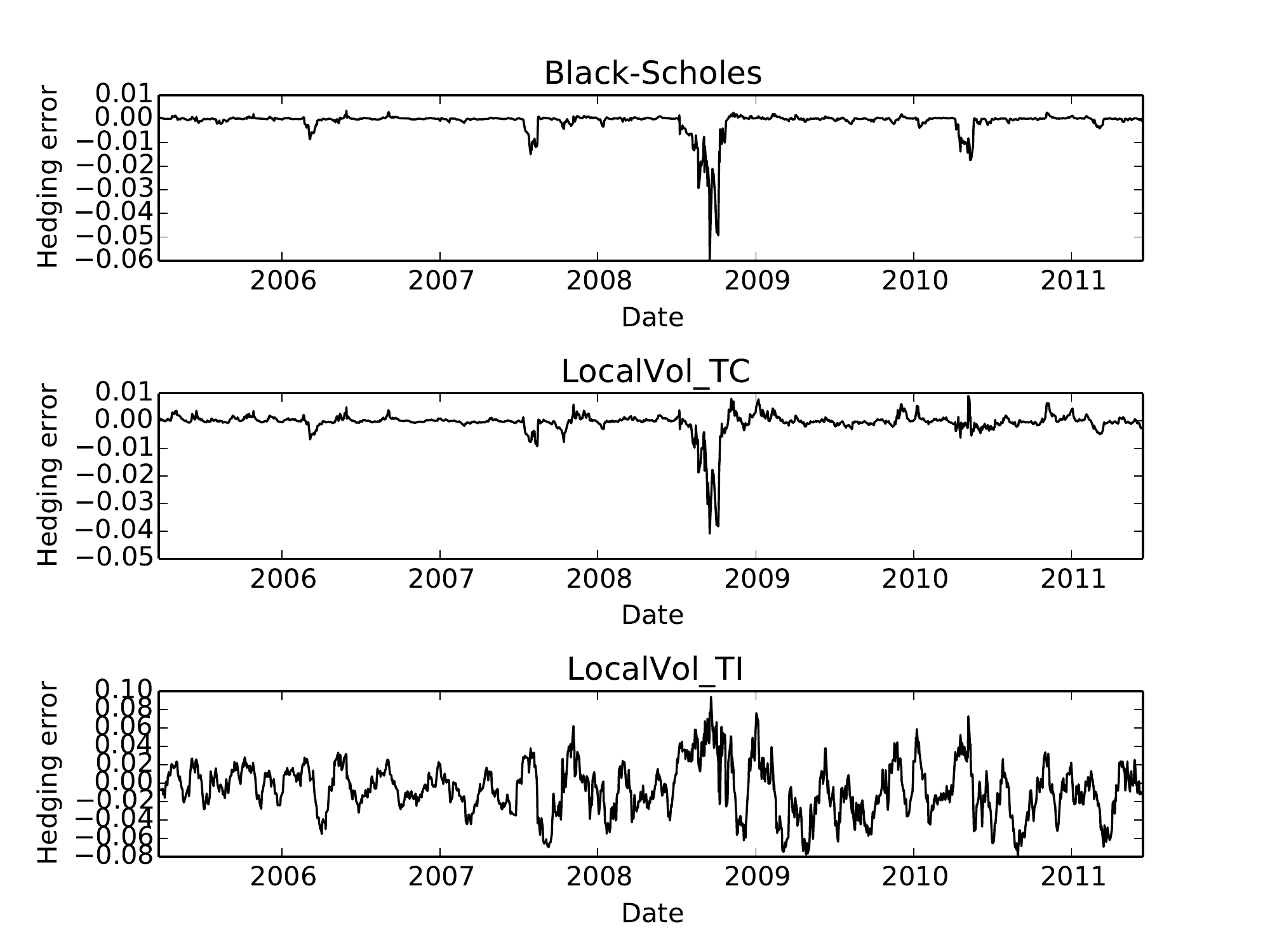}}
  \caption{Delta hedging errors for (10$\Delta$Put) calls with one
    month maturity.}
  \label{figs:hedging_errors_one_month_Put10D}
\end{figure}

\begin{figure}
  \centering \subfloat[][Histogram]
  {\label{fig:hist_one_month_Put25D}\includegraphics[width=0.52\textwidth]{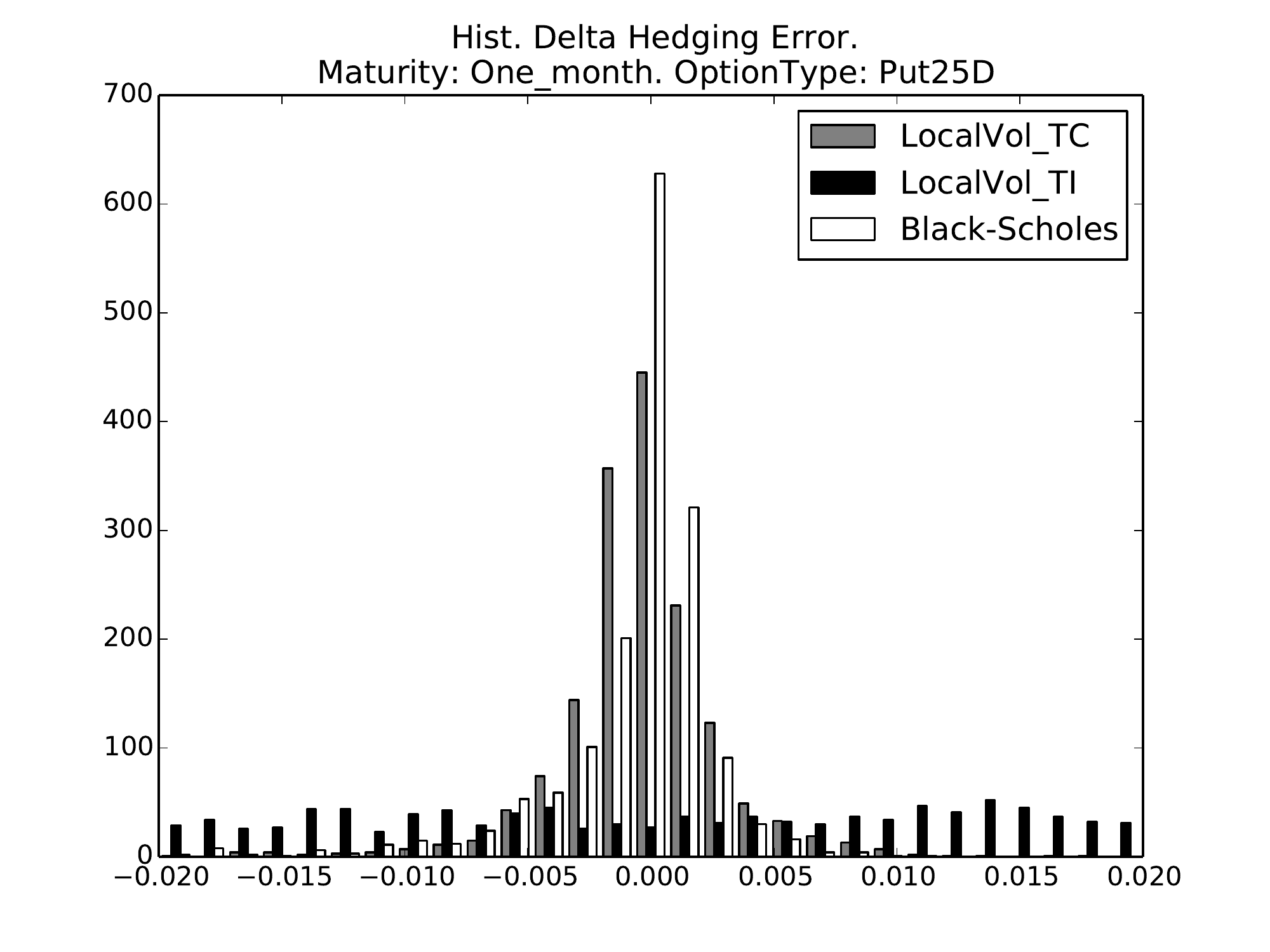}}
  \centering \subfloat[][Hedging error time series]
  {\label{fig:TS_one_month_Put25D}\includegraphics[width=0.52\textwidth]{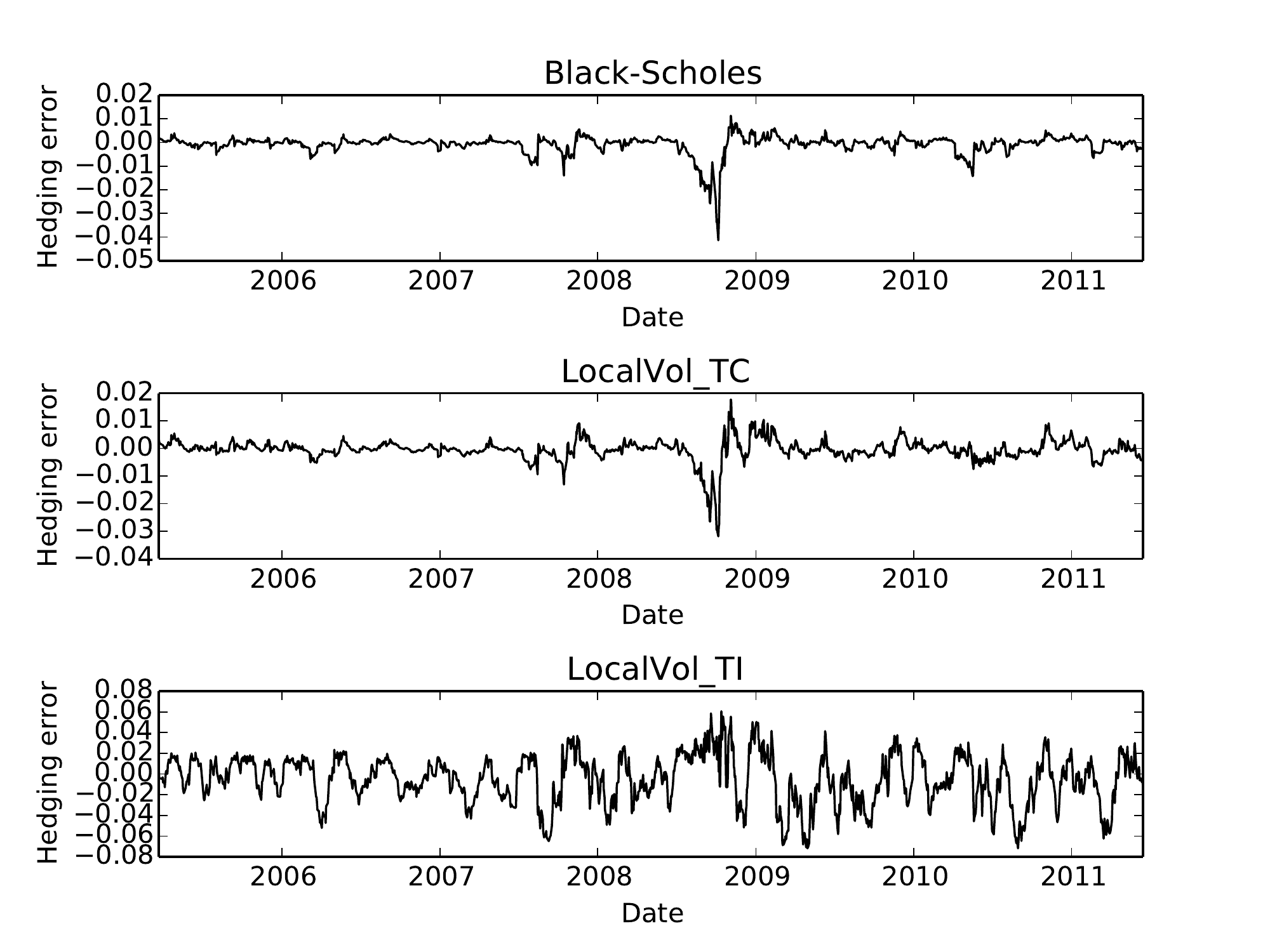}}
  \caption{Delta hedging errors for (25$\Delta$Put) calls with one
    month maturity.}
  \label{figs:hedging_errors_one_month_Put25D}
\end{figure}

\begin{figure}
  \centering \subfloat[][Histogram]
  {\label{fig:hist_one_month_ATM}\includegraphics[width=0.52\textwidth]{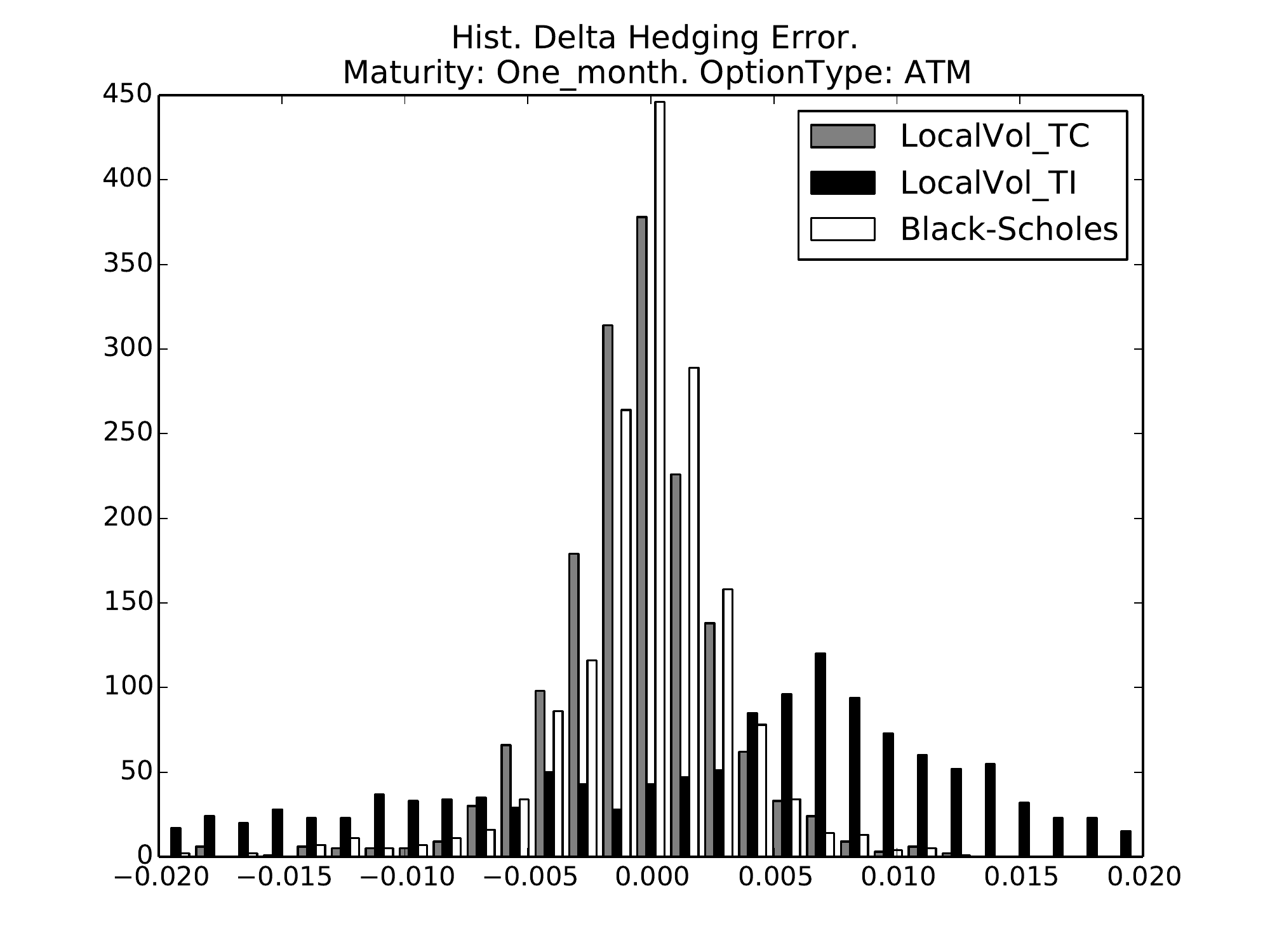}}
  \centering \subfloat[][Hedging error time series]
  {\label{fig:TS_one_month_ATM}\includegraphics[width=0.52\textwidth]{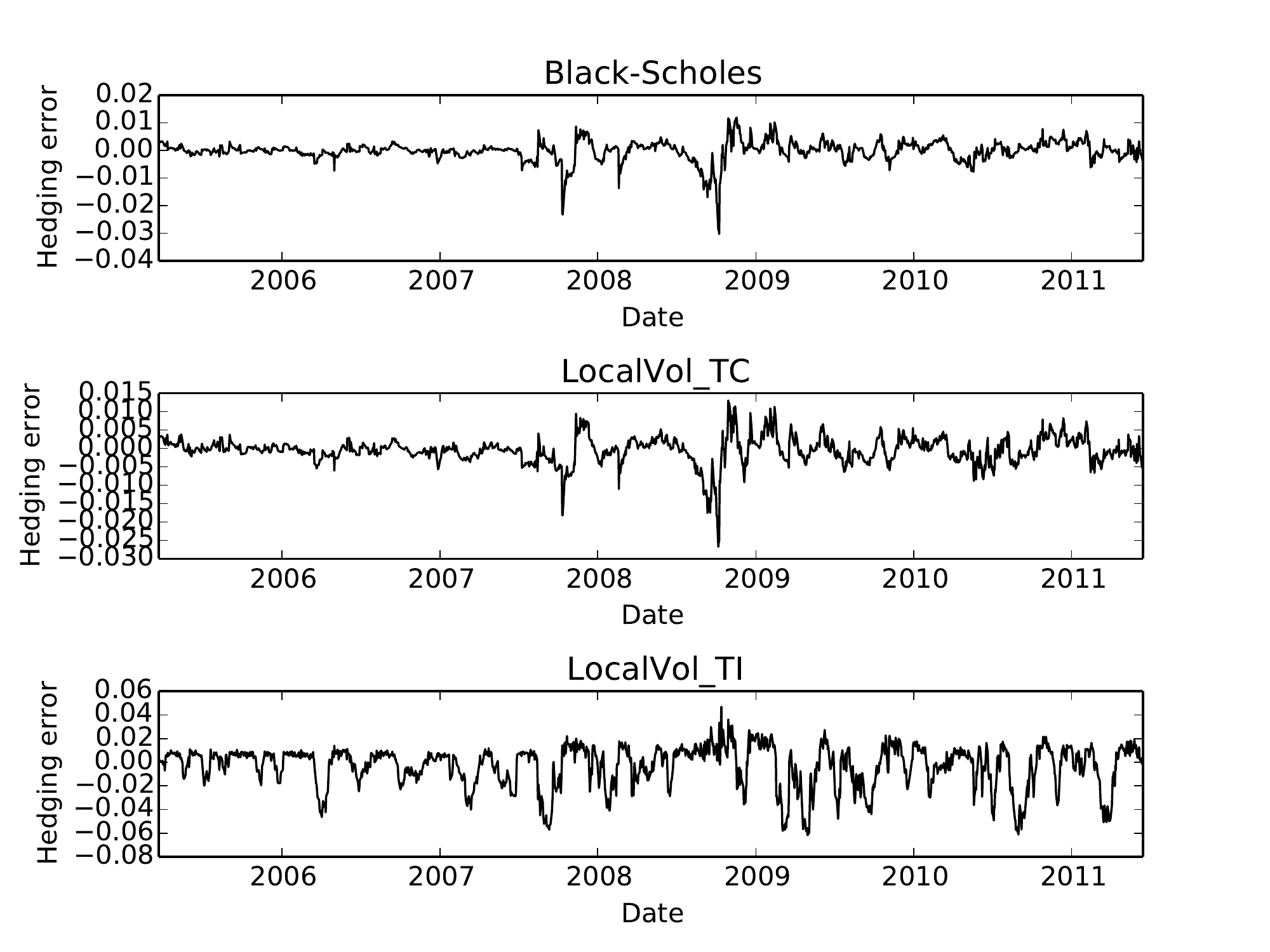}}
  \caption{Delta hedging errors for ATM calls with one
    month maturity.}
  \label{figs:hedging_errors_one_month_ATM}
\end{figure}

\begin{figure}
  \centering \subfloat[][Histogram]
  {\label{fig:hist_one_month_Call25D}\includegraphics[width=0.52\textwidth]{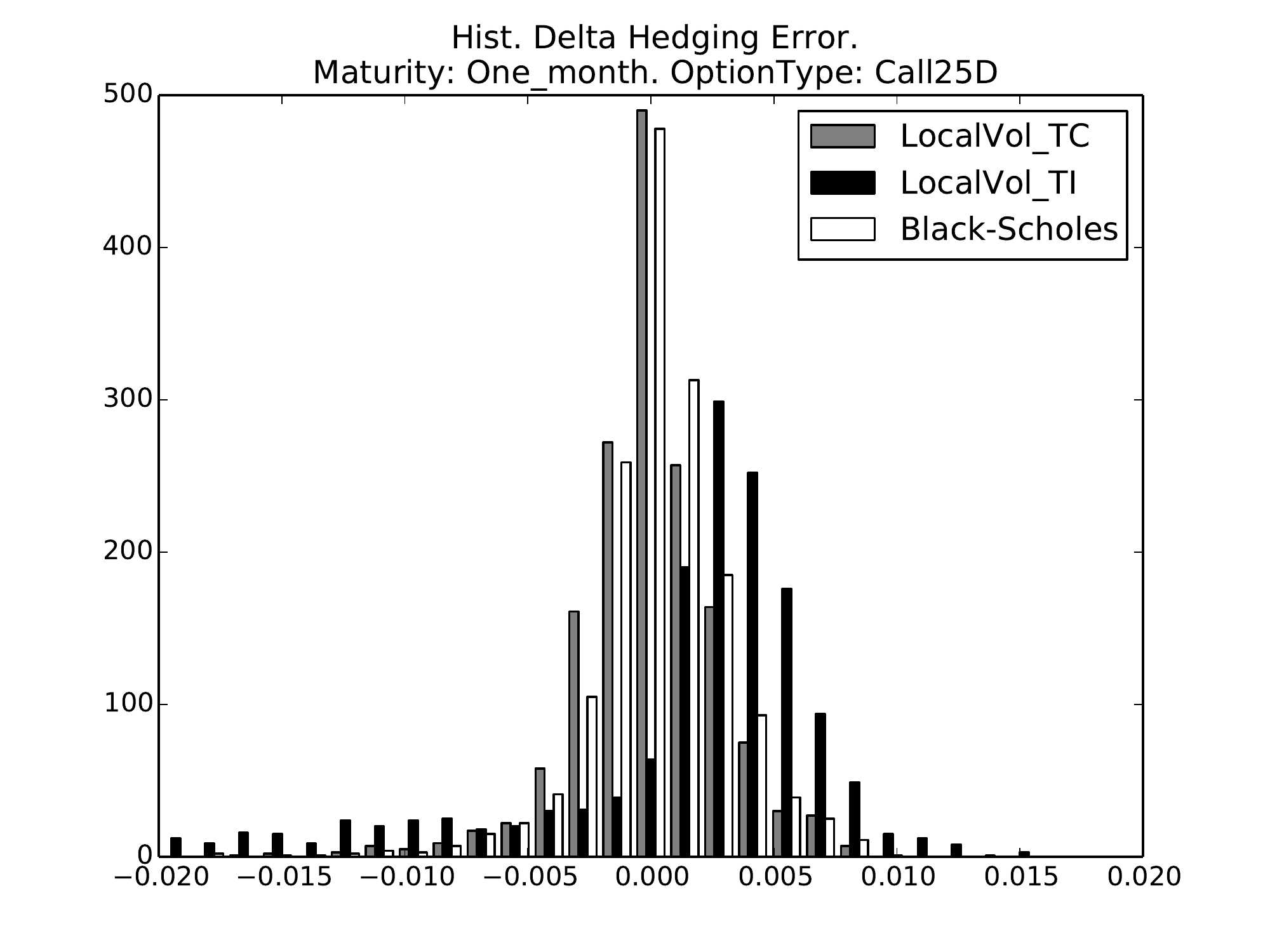}}
  \centering \subfloat[][Hedging error time series]
  {\label{fig:TS_one_month_Call25D}\includegraphics[width=0.52\textwidth]{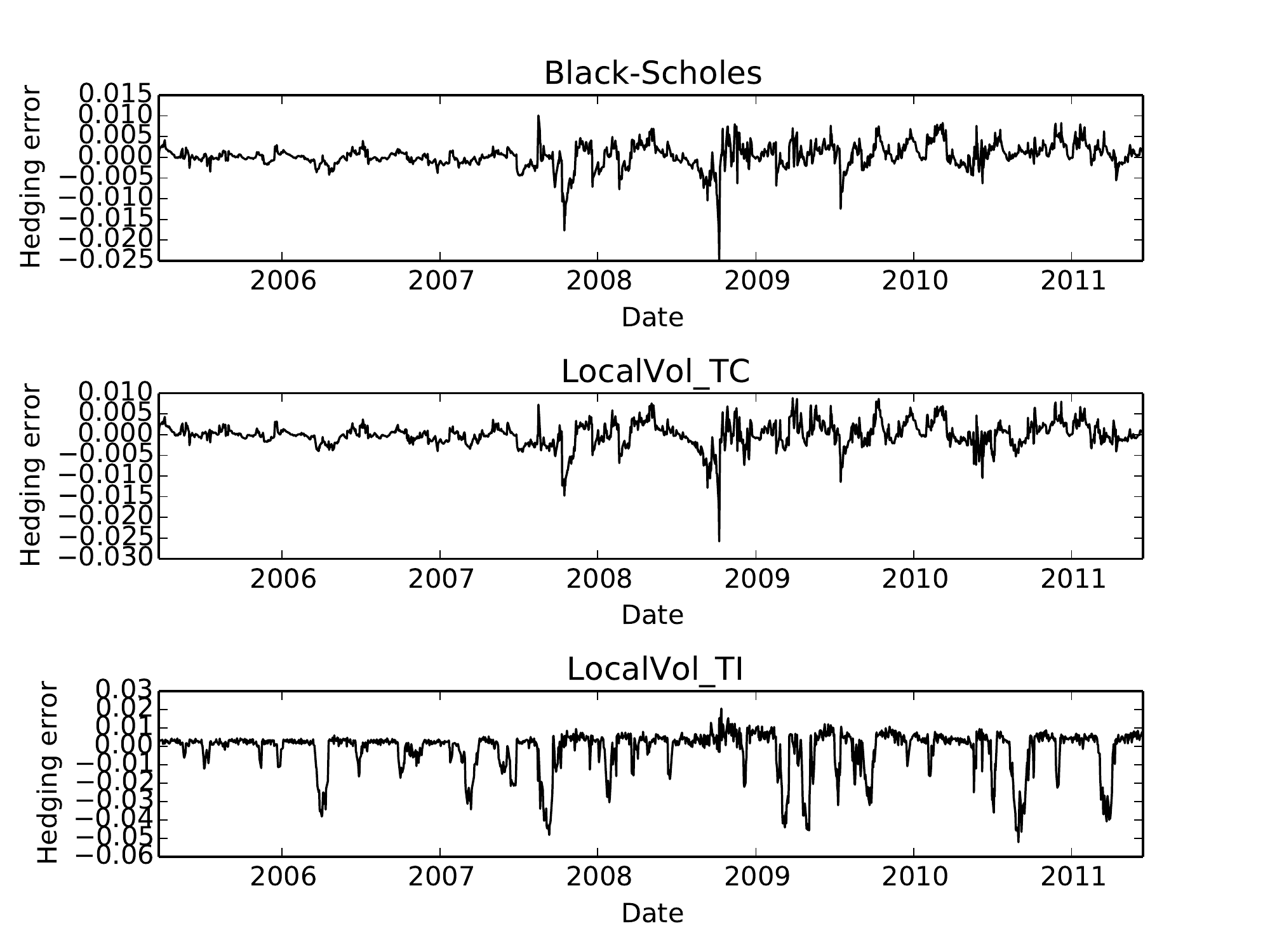}}
  \caption{Delta hedging errors for (25$\Delta$Call) calls with one
    month maturity.}
  \label{figs:hedging_errors_one_month_Call25D}
\end{figure}

\begin{figure}
  \centering \subfloat[][Histogram]
  {\label{fig:hist_one_month_Call10D}\includegraphics[width=0.52\textwidth]{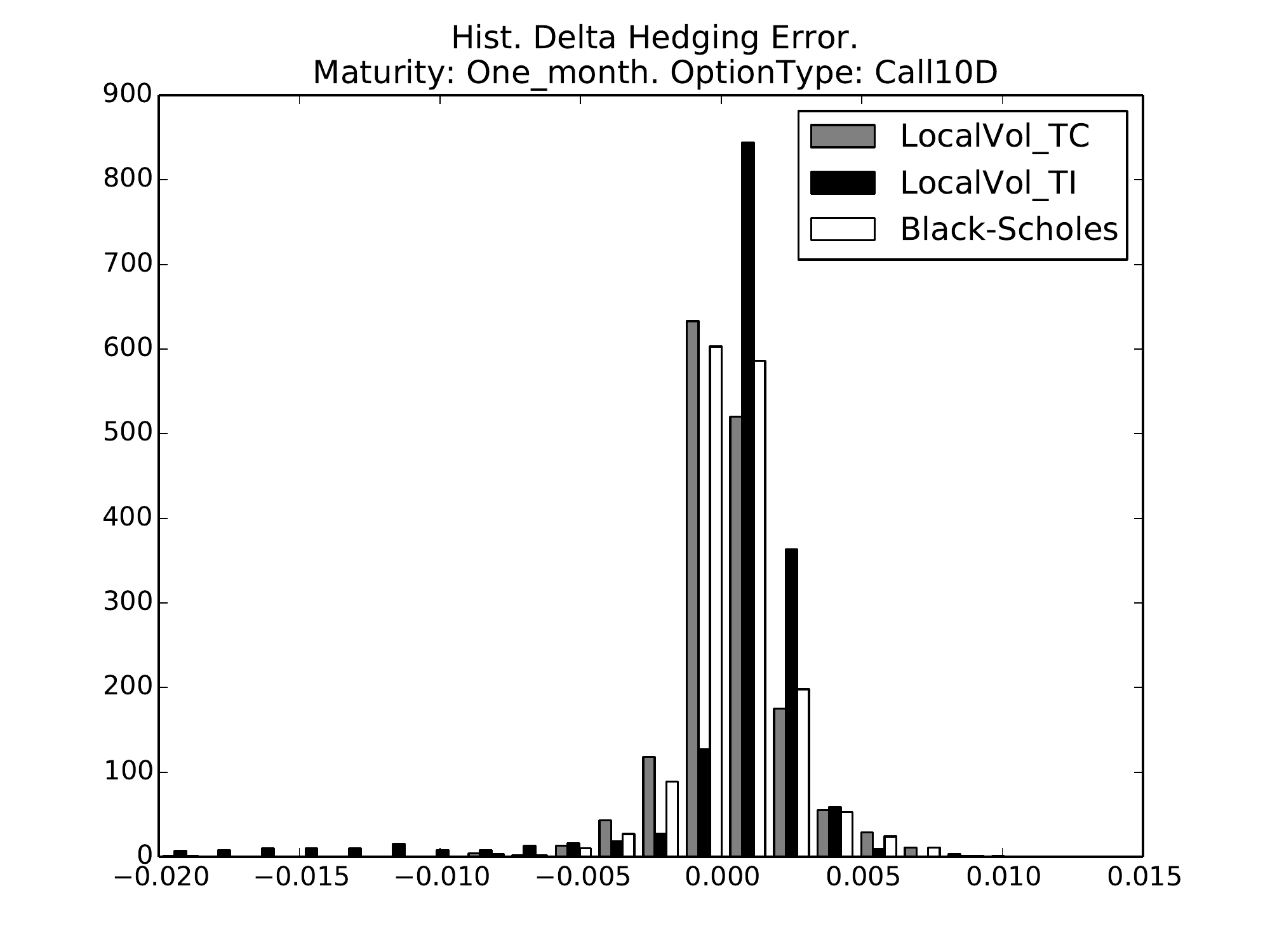}}
  \centering \subfloat[][Hedging error time series]
  {\label{fig:TS_one_month_Call10D}\includegraphics[width=0.52\textwidth]{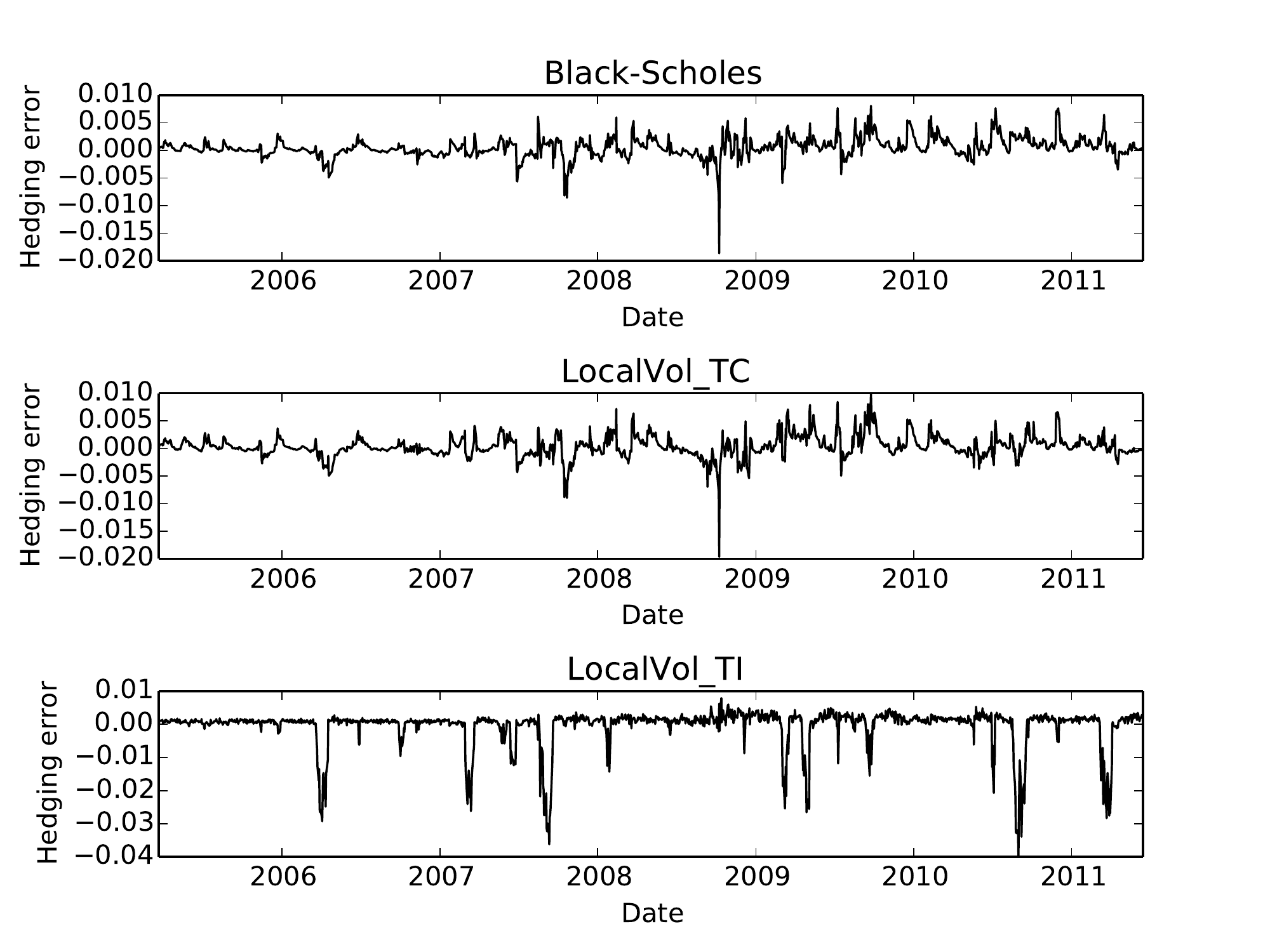}}
  \caption{Delta hedging errors for (10$\Delta$Call) calls with one
    month maturity.}
  \label{figs:hedging_errors_one_month_Call10D}
\end{figure}

\begin{table}[htbp]
\centering
\begin{tabular}{rrrr}
\hline
\textbf{Delta} & \textbf{Model} & \textbf{Mean} & \textbf{Std. dev.}\\
\hline
\(10\Delta\)Put & Black-Scholes & -0.0004 & 0.0024\\
\(10\Delta\)Put & LocalVol\_TC & -0.0001 & 0.0023\\
\(10\Delta\)Put & LocalVol\_TI & -0.001 & 0.0129\\
\(25\Delta\)Put & Black-Scholes & -0.0004 & 0.0029\\
\(25\Delta\)Put & LocalVol\_TC & -0.0001 & 0.0028\\
\(25\Delta\)Put & LocalVol\_TI & -0.0009 & 0.0108\\
ATM & Black-Scholes & -0.0 & 0.0028\\
ATM & LocalVol\_TC & -0.0003 & 0.0028\\
ATM & LocalVol\_TI & -0.0008 & 0.0078\\
\(25\Delta\)Call & Black-Scholes & 0.0003 & 0.0023\\
\(25\Delta\)Call & LocalVol\_TC & 0.0002 & 0.0024\\
\(25\Delta\)Call & LocalVol\_TI & -0.0 & 0.0046\\
\(10\Delta\)Call & Black-Scholes & 0.0003 & 0.0016\\
\(10\Delta\)Call & LocalVol\_TC & 0.0003 & 0.0016\\
\(10\Delta\)Call & LocalVol\_TI & 0.0003 & 0.002\\
\hline
\end{tabular}
\caption{Mean and standard deviation of hedging errors for each model
  under consideration for one week maturity.}
\label{tab:hedgeErrorTable_oneweek}
\end{table}

\begin{table}[htbp]
\centering
\begin{tabular}{rrrr}
\hline
\textbf{Delta} & \textbf{Model} & \textbf{Mean} & \textbf{Std. dev.}\\
\hline
\(10\Delta\)Put & Black-Scholes & -0.0012 & 0.005\\
\(10\Delta\)Put & LocalVol\_TC & -0.0006 & 0.004\\
\(10\Delta\)Put & LocalVol\_TI & -0.0063 & 0.0275\\
\(25\Delta\)Put & Black-Scholes & -0.0008 & 0.0041\\
\(25\Delta\)Put & LocalVol\_TC & -0.0005 & 0.0039\\
\(25\Delta\)Put & LocalVol\_TI & -0.0051 & 0.0236\\
ATM & Black-Scholes & -0.0002 & 0.0036\\
ATM & LocalVol\_TC & -0.0006 & 0.0036\\
ATM & LocalVol\_TI & -0.0035 & 0.0177\\
\(25\Delta\)Call & Black-Scholes & 0.0003 & 0.0029\\
\(25\Delta\)Call & LocalVol\_TC & -0.0001 & 0.0029\\
\(25\Delta\)Call & LocalVol\_TI & -0.0016 & 0.0114\\
\(10\Delta\)Call & Black-Scholes & 0.0005 & 0.0018\\
\(10\Delta\)Call & LocalVol\_TC & 0.0003 & 0.002\\
\(10\Delta\)Call & LocalVol\_TI & -0.0005 & 0.0061\\
\hline
\end{tabular}
\caption{Mean and standard deviation of hedging errors for each model
  under consideration for one month maturity.}
\label{tab:hedgingErrorTable_onemonth}
\end{table}

\section{Results}
\label{sec:results}

Figures 1 to 10 depict histograms of delta hedging errors computed
from the historical data under the frameworks of Black-Scholes and
local volatility. Within each histogram \emph{LocalVol\_TC} and
\emph{LocalVol\_TI} denote the theoretically correct delta and sticky
delta approaches, respectively. Sample means and standard deviations
for these hedging errors are summarised in Tables
\ref{tab:hedgeErrorTable_oneweek} and
\ref{tab:hedgingErrorTable_onemonth}.

We note that for in and at-the-money options (figures 1 to 3 for one
week maturity and figures 6 to 8 for one month maturity), that
the local volatility model with sticky delta performs significantly
worse to the other two methods. It is only with deep in-the-money
options (figure \ref{fig:hist_one_week_Call10D}) that sticky delta
local volatility exhibits hedging errors better than Black-Scholes.

\section{Conclusion}
Using delta hedging as the criterion to measure the effectiveness of a
market model, our results show that Black-Scholes is no worse than the
local volatility model. In fact the Black-Scholes model performs
significantly better than sticky delta local volatility, particularly
for in and at-the-money options. The theoretically correct delta local
volatility model gives hedging errors which are not too far from that
of Black-Scholes and stick delta local volatility performs noticeably
worse than the other models except for the case of deep out of the
money options.

Further avenues of research include performing these empirical tests on
other FX pairs and also incorporating other hedges such as vega. Also
the framework can be used to validate/compare other models such as
stochastic volatility, local stochastic volatility etc. It will also
be of interest to determine hedging errors for exotic options such as
barrier options.


\section*{Acknowledgment}

The authors would like to thank Igor Geninson and Rod Lewis from
Commonwealth bank of Australia Global Markets for their assistance as
well as Xiaolin Luo from CSIRO Mathematics, Informatics and Statistics.



%

\bibliographystyle{IEEEtran}
\bibliography{references}


\end{document}